\documentclass[prd,12pt,nofootinbib,preprint]{revtex4}

\usepackage[T1]{fontenc}
\usepackage{amsmath,amssymb}
\usepackage{relsize}
\usepackage{epsfig}
\usepackage{url}
\usepackage{subfigure}
\usepackage{graphicx}
\usepackage{grffile}
\usepackage[usenames,dvipsnames]{color}
\usepackage{slashed}
\usepackage{array}
\usepackage{multirow}
\usepackage[colorlinks,citecolor=blue]{hyperref}
\usepackage{color}
\usepackage{cancel}
\usepackage{gensymb}
\usepackage{soul}

\def \met{\slashed{E}_T }

\def\a {\alpha}
\def\b {\beta}
\def\g {\gamma}

\def\l {\lambda}

\def\bar {\overline}

\def\be {\begin{equation}}
\def\ee {\end{equation}}
\def\beq {\begin{equation}}
\def\eeq {\end{equation}}
\def\bea {\begin{eqnarray}}
\def\eea {\end{eqnarray}}

\newcommand{\besub}{\begin{subequations}}
\newcommand{\eesub}{\end{subequations}}

\def\g{\Gamma}

\def\beq{\begin{equation}}
\def\eeq{\end{equation}}
\def\barr{\begin{array}}
\def\earr{\end{array}}

\begin{document}
\title{Muon $g-2$ and $W$-mass in a framework of colored scalars: an LHC perspective}

\author{Nabarun Chakrabarty}
\email{nabarunc@iitk.ac.in}
\affiliation{Department of Physics, Indian Institute of Technology Kanpur, Kanpur, Uttar Pradesh-208016, India} 

\author{Indrani Chakraborty}
\email{indranic@iitk.ac.in, indrani300888@gmail.com}
\affiliation{Department of Physics, Indian Institute of Technology Kanpur, Kanpur, Uttar Pradesh-208016, India} 

\author{Dilip Kumar Ghosh}
\email{tpdkg@iacs.res.in}
\affiliation{School of Physical Sciences, Indian Association for the Cultivation of Science,\\ 2A $\&$ 2B, Raja S.C. Mullick Road, Jadavpur, Kolkata 700032, India }

\author{Gourab Saha}
\email{gourab.saha@saha.ac.in}
\affiliation{Saha Institute of Nuclear Physics, 1/AF Bidhan Nagar, Kolkata 700064, India}

\begin{abstract} 
A color octet isodoublet can have esoteric origins and it complies with minimal flavour violation. In this study, we take a scenario where the well known Type-X Two-Higgs doublet model is augmented with a color octet isodoublet. We shed light on how such a setup can predict the recently observed value for the $W$-boson mass. We also evaluate the two-loop Barr-Zee contributions to muon $g-2$ stemming from the colored scalars. The parameter space compatible with the observed muon $g-2$ gets relaxed \emph{w.r.t.} what it is in the pure Type-X 2HDM by virtue of the contribution from the colored scalars. The extended parameter region therefore successfully accounts for both the $W$-mass and muon $g-2$ anomalies successfully. Finally, a collider signature leading to $\tau^+ \tau^- b \bar{b}$ final state is explored at the 14 TeV LHC using both cut-based and multivariate techniques. Such a signal can confirm the existence of both colorless as well colored scalars that are introduced by this framework.
\end{abstract} 
\maketitle

\section{Introduction} 
The particle spectrum of the Standard Model (SM) is deemed complete following the discovery of a Higgs boson~\cite{Chatrchyan:2012xdj,Aad:2012tfa} at the Large Hadron Collider (LHC). Additionally, the interaction strengths of the Higgs with the SM fermions and gauge bosons are in good agreement with the SM predictions. Despite such triumph of the SM, some longstanding issues on both theoretical and experimental fronts have long been advocating additional dynamics beyond the SM (BSM). Such issues include a non-zero neutrino mass, the existence of dark matter (DM), the observed imbalance between matter and antimatter in the universe, and, the instability (or metastability) of the electroweak (EW) vacuum~\cite{EliasMiro:2011aa,Bezrukov:2012sa,Degrassi:2012ry,Buttazzo:2013uya} in the SM. Interestingly, extensions of the SM Higgs sector can serve as powerful prototypes of BSM physics that can potentially solve the aforesaid issues. 

Apart from the longstanding issues, some recent experimental observations have thrown fresh insight on as to what could be the nature of some hitherto additional dynamics beyond the SM. One example is the recently reported value of the mass of the $W$-boson by the CDF collaboration, that is deviated \emph{w.r.t.} the SM prediction~\cite{Blum:2013xva,RBC:2018dos,Keshavarzi:2018mgv,Davier:2019can,Aoyama:2020ynm,Colangelo:2018mtw,Hoferichter:2019mqg,Melnikov:2003xd,Hoferichter:2018kwz,Blum:2019ugy,ParticleDataGroup:2020ssz} by 7.2$\sigma$. That is,
\bea
M^{\text{CDF}}_W &=&  80.4335~\text{GeV} \pm 6.4~\text{MeV} (stat) \pm 6.9~\text{MeV} (sys).
\eea

The origin of this deviation is suspected to be some New Physics (NP). The second experimental result is the reporting of an excess in the anomalous magnetic moment of the muon by FNAL~\cite{Muong-2:2021ojo,Muong-2:2021vma}, thereby concurring with the earlier result by BNL~\cite{Muong-2:2006rrc}. The combined result is quoted as
\bea
\Delta a_\mu = (2.51 \pm 0.59) \times 10^{-9}.
\eea
A Two-Higgs doublet model (2HDM)~\cite{Branco:2011iw,Deshpande:1977rw} with a Type-X texture for Yukawa interactions has been long known to address the muon $g-2$ excess. The scalar sector of a 2HDM comprises the CP-even neutral scalars $h,H$, the CP-odd neutral scalar $A$, and a singly charged scalar $H^+$. Here, $h$ denotes the SM-like Higgs with mass 125 GeV. The vacuum expectation values of two doublets are $v_1$ and $v_2$ with tan$\beta = \frac{v_2}{v_1}$. Demanding invariance under a $\mathbb{Z}_2$ symmetry with the aim of annulling flavour changing neutral currents (FCNCs) leads to several variants of the 2HDM a particular kind of which is the Type-X. This variant features enhanced leptonic Yukawas with $H$ and $A$
and an sizeable contributions to muon $g-2$ are introduced via two-loop Barr-Zee (BZ) amplitudes. A resolution of the anomaly thus becomes possible for a light $A$ ($M_A \lesssim$ 100 GeV) and high tan$\beta$ ($\gtrsim 20$)~\cite{Broggio:2014mna,Cao:2009as,Wang:2014sda,Ilisie:2015tra,Abe:2015oca,Chun:2016hzs,Cherchiglia:2016eui,Dey:2021pyn}. The 2HDM framework can also accommodate $M_W^{\text{CDF}}$ \cite{Lee:2022gyf,Song:2022xts,Bahl:2022xzi,Babu:2022pdn,Ahn:2022xax,Han:2022juu,Arcadi:2022dmt,Ghorbani:2022vtv,Benbrik:2022dja,Botella:2022rte,Kim:2022xuo,Kim:2022hvh,Appelquist:2022qgl,Atkinson:2022qnl,Hessenberger:2022tcx,Kim:2022nmm,Arco:2022jrt,Kang:2022mdy,Jung:2022prq}. However, 
stringent constraints coming from lepton flavour universality in $\tau$ decays restricts large tan$\beta$. Also, recent LHC searches for $h \to AA \to 4\tau, 2\tau 2\mu$~\cite{CMS:2018qvj} channels rules out
a large $h \to A A$ branching ratio. Such experimental results restrict to a great extent the parameter space in the Type-X that favours an explanation of muon $g-2$.
A possible way to relax the parameter space is to introduce additional scalar degrees of freedom so that additional BZ amplitudes are induced.

An interesting extension of the SM involves a scalar multiplet transforming as (8,2,1/2)~\cite{Manohar:2006ga} under the SM gauge group. Such a scenario is motivated by minimal flavour violation (MFV). It assumes all breaking of the underlying approximate flavour symmetry
of the SM is proportional to the up- or down-quark Yukawa matrices. And it has been shown
in \cite{Manohar:2006ga} that the only scalar representations under the SM gauge group complying with
MFV are (\textbf{1,2}, 1/2 ) and (\textbf{8,2}, 1/2 ). The colored scalars emerging from this multiplet are the CP-even $S_R$, the CP-odd $S_I$ and the singly charged $S^+$.  In addition, a color-octet can also stem from Grand Unification~\cite{Popov:2005wz,Dorsner:2007fy,FileviezPerez:2008ib,Perez:2008ry}, topcolor models~\cite{Hill:1991at} and extra dimensional scenarios~\cite{Dobrescu:2007xf,Dobrescu:2007yp}. Important phenomenological consequences of such a construct were studied in \cite{Carpenter:2011yj,Enkhbat:2011qz,Arnold:2011ra,Kribs:2012kz,Cao:2013wqa,Ding:2016ldt,Cao:2015twy,Gerbush:2007fe}. In fact,
a scenario augmenting a 2HDM with a color-octet isodoublet has also been discussed in \cite{Cheng:2016tlc,Cheng:2017tbn}. The Type-I and Type-II variants were employed here. Important exclusion limits on such a framework were deduced in \cite{Miralles:2019uzg} and 
the radiatively generated $H^+ W^- Z(\gamma)$ vertex was studied in \cite{Chakrabarty:2020msl}.

In this work, we extend the Type-X 2HDM by a color-octet iso-doublet. Taking into account the various constraints on this setup, we first identify the parameter region that accounts for $M^{\text{CDF}}_W$. We subsequently demonstrate how the parameter space accommodating $\Delta a_\mu$ expands \emph{w.r.t.} the pure Type-X on account of the additional BZ amplitudes stemming from the colored scalars. Thus, the given framework is shown to address the two anomalies simultaneously. We also propose a collider signal $p p \to S_R \to S_I A,~S_I \to b \bar{b},~A \to \tau^+ \tau^-$ for a hadron collider. Such a  final state gives information about both the colorless and colored scalars involved in the cascade. In addition to the conventional cut-based methods, we plan to also use the more modern multivariate techniques for the analysis.

The study is organised as follows. We introduce the Type-X 2HDM plus color-octet framework in section \ref{model}. In section \ref{constraints}, we list the important constraints on this model from theory and experiments.
The resolution of the $W$-mass and muon $g-2$ anomalies in detailed in section \ref{anomalies}. A detailed analysis of the proposed LHC signature is presented in \ref{collider} employing both cut-based as well as multivariate techniques. Finally, the study is concluded in \ref{conclusions}. Important formulae are given in the Appendix.

\section{The Type-X + color octet framework}\label{model}
The scalar sector of the framework consists of two color-singlet $SU(2)_L$ scalar doublets $\Phi_{1,2}$ and one color-octet 
$SU(2)_L$ scalar $S$. The multiplets are parametrised as:
\begin{eqnarray}
\Phi_i = \begin{pmatrix}
\phi_i^+ \\
\frac{1}{\sqrt{2}} (v_i + h_i + i z_i)
\end{pmatrix} , (i = 1,2) ,~ 
S = \begin{pmatrix}
S^+ \\
\frac{1}{\sqrt{2}} (S_R + i S_I)
\end{pmatrix}.
\end{eqnarray} 
The electroweak gauge group $SU(2)_L \times U(1)_Y$ is spontaneously broken to $U(1)_Q$ when $\Phi_i$ receives a 
vacuum expectation value (VEV) $v_i$ with $v^2 = v_1^2 + v_2^2 = (246 ~{\rm GeV})^2$. That the multiplet $S$ receives no VEV averts a spontaneous breakdown of $SU(3)_c$. 

The most generic scalar potential consistent with the gauge symmetry consists of a part containing the interactions among 
$\Phi_{1,2}$ only ($V_a(\Phi_{1},\Phi_{2})$), a part containing only $S$ ($V_b(S)$) and  
a part containing the interactions among all $\Phi_{1,2},S$ ($V_c(\Phi_{1},\Phi_{2},S)$). 
The scalar potential therefore looks like~\cite{Cheng:2016tlc}
\bea
V (\Phi_{1},\Phi_{2},S) &=& V_a (\Phi_{1},\Phi_{2}) + V_b(S) + V_c(\Phi_{1},\Phi_{2},S), 
\eea
where,
\bea
V_a (\Phi_{1},\Phi_{2})&=& m_{11}^2 \Phi_1^\dag \Phi_1 + m_{22}^2 \Phi_2^\dag \Phi_2 - m_{12}^2 \left( \Phi_1^\dag \Phi_2 + \Phi_2^\dag \Phi_1 \right) \nonumber \\
&& +  \frac{\lambda_1}{2} \left( \Phi_1^\dag \Phi_1 \right)^2 + \frac{\lambda_2}{2} \left( \Phi_2^\dag \Phi_2 \right)^2 
+ \lambda_3 \left( \Phi_1^\dag \Phi_1 \right) \left( \Phi_2^\dag \Phi_2 \right) + \lambda_4 \left( \Phi_1^\dag \Phi_2 \right) \left( \Phi_2^\dag \Phi_1 \right) \nonumber \\
&& +  \left[ \frac{\lambda_5}{2} \left( \Phi_1^\dag \Phi_2 \right)^2  + \lambda_6 \left( \Phi_1^\dag \Phi_1 \right) \left( \Phi_1^\dag \Phi_2 \right)  + \lambda_7 \left( \Phi_2^\dag \Phi_2 \right) \left( \Phi_1^\dag \Phi_2 \right)+ {\rm H.c.}\right],
\label{pot-1}
\eea
\bea
V_b(S) &=& 2m_S^2 {\rm Tr}S^{\dag i}S_i + \mu_1 {\rm Tr}S^{\dag i}S_i S^{\dag j}S_j + \mu_2 {\rm Tr}
S^{\dag i}S_j S^{\dag j}S_i + \mu_3 {\rm Tr} S^{\dag i}S_i {\rm Tr}S^{\dag j} S_j\nonumber\\
& +& \mu_4 {\rm Tr}S^{\dag i}S_j {\rm Tr}S^{\dag j}S_i + \mu_5 {\rm Tr}S_i S_j{\rm Tr}
S^{\dag i}S^{\dag j} + \mu_6 {\rm Tr}S_i S_j S^{\dag j}S^{\dag i} \,,
\label{pot-2}
\eea

\bea
V_c(\Phi_{1},\Phi_{2},S) &=& \nu_1 \Phi_1^{\dag i}\Phi_{1i}{\rm Tr}S^{\dag j}S_j + \nu_2 \Phi_1^{\dag i}\Phi_{1j}
{\rm Tr}S^{\dag j}S_i\nonumber\\
& +& \left( \nu_3 \Phi_1^{\dag i}\Phi_1^{\dag j}{\rm Tr}S_i S_j + \nu_4 \Phi_1^{\dag i}{\rm Tr}
S^{\dag j}S_j S_i + \nu_5 \Phi_1^{\dag i}{\rm Tr}S^{\dag j}S_i S_j + {\rm h.c.} \right) \nonumber \\
&+& \omega_1 \Phi_2^{\dag i}\Phi_{2i}{\rm Tr}S^{\dag j}S_j + \omega_2 \Phi_2^{\dag i}\Phi_{2j}
{\rm Tr}S^{\dag j}S_i\nonumber\\
& +& \left( \omega_3 \Phi_2^{\dag i}\Phi_2^{\dag j}{\rm Tr}S_i S_j + \omega_4 \Phi_2^{\dag i}{\rm Tr}
S^{\dag j}S_j S_i + \omega_5 \Phi_2^{\dag i}{\rm Tr}S^{\dag j}S_i S_j + {\rm h.c.} \right) \nonumber\\
&+& \kappa_1 \Phi_1^{\dag i}\Phi_{2i}{\rm Tr}S^{\dag j}S_j +
\kappa_2 \Phi_1^{\dag i}\Phi_{2j}{\rm Tr}S^{\dag j}S_i + \kappa_3 \Phi_1^{\dag i}\Phi_2^{\dag j}{\rm Tr}S_j S_i,
+ {\rm h.c.}
\label{pot-3}
\eea

Here,  $i,j$ denote the fundamental $SU(2)$ indices. One can define $S_i = S_i^B T^B$ ($T^B$ being the $SU(3)$ generators and $'B'$ being the $SU(3)$ adjoint index) and the traces in Eq.(\ref{pot-2}) and Eq.(\ref{pot-3}) are taken over the color indices. We mention here that we do not impose some ad-hoc discrete symmetry to restrict the scalar potential. Rather, we are guided purely by MFV \cite{Manohar:2006ga}. One clearly identifies $V_a(\Phi_{1},\Phi_{2})$ with the generic scalar potential of two Higgs doublet model (2HDM). An important 2HDM parameter is
$\tan \beta = \frac{v_2}{v_1}$. We take the VEVs and all model parameters to be real in order to avoid $\text{CP}$-violation. The scalar spectrum expectedly consists of both color-singlet as well as color-octet particles. 

The color-singlet scalar mass spectrum comprising the $\text{CP}$-even $h,H$, a $\text{CP}$-odd 
$A$ and a charged Higgs $H^+$, coincides with that of a 2HDM. Of these, $h$ is identified with the discovered scalar with mass 125 GeV.
The expressions of the physical masses belonging to the particles in the colorless counterpart in terms of the couplings and mixing angles $\beta$ and $\alpha$\footnote{$\alpha $ is the mixing angle in the $\text{CP}$-even sector.} could be found in \cite{Branco:2011iw}. On the other hand, the masses of the neutral ($S_R,S_I$) and charged mass eigenstate ($S^+$) of the color-octet can be expressed in terms of the quartic couplings 
$\omega_i, \kappa_i, \nu_i$ and mixing angle $\beta$ as \cite{Cheng:2016tlc}:
\besub
\bea
M_{S_R}^2 &=& m_S^2 + \frac{1}{4} v^2 (\cos ^2 \beta  (\nu_1 + \nu_2 + 2 \nu_3)+\sin 2 \beta  (\kappa_1 + \kappa_2 + \kappa_3) \nonumber \\
&& + \sin ^2 \beta  (\omega_1 + \omega_2 + 2 \omega_3)) \,, \\
M_{S_I}^2 &=& m_S^2 + \frac{1}{4} v^2 (\cos ^2 \beta  (\nu_1 + \nu_2 - 2 \nu_3)+\sin 2 \beta  (\kappa_1 + \kappa_2 - \kappa_3) \nonumber \\
&& + \sin ^2 \beta  (\omega_1 + \omega_2 - 2 \omega_3)) \,, \\
M_{S^+}^2 &=& m_S^2 + \frac{1}{4} v^2 (\nu_1 \cos ^2 \beta + \kappa_1 \sin 2 \beta + \omega_1 \sin ^2 \beta ).
\eea
\eesub

The Yukawa interactions in this framework are discussed next. For the interactions involving $\phi_1$ and $\phi_2$, we adopt the Type-X 2HDM Lagrangian. Here, the quarks get their masses from $\phi_2$ and the leptons, from $\phi_1$. That is,
\bea
-\mathcal{L}^{\text{2HDM}}_Y &=& \Big[ y_u \bar{Q_L} \tilde{\phi}_2 u_R + y_d \bar{Q_L} \phi_2 d_R + y_\ell \bar{Q_L} \phi_1 \ell_R \Big] + \text{h.c.}
\eea
The lepton Yukawa interactions in terms of the physical scalars then becomes
\bea
\mathcal{L}^\text{2HDM}_Y &=& \sum_{\ell=e,\mu,\tau} \frac{m_\ell}{v} \bigg(\xi_\ell^h h \bar{\ell} \ell + \xi_\ell^H H \bar{\ell} \ell - i \xi_\ell^A A \bar{\ell} \gamma_5 \ell + \Big[ \sqrt{2} \xi^A_\ell H^+ \bar{\nu_\ell} P_R \ell + \text{h.c.} \Big] \bigg).
\eea
The various $\xi_\ell$ factors are tabulated in the Appendix.

The Yukawa interactions of the colored scalars can be expressed as \cite{Manohar:2006ga}
\bea
-\mathcal{L}^{\text{col. oct.}}_Y &=& \sum_{p,q=1,2,3} \Big[Y^{pq}_u~\bar{Q_{Lp}} \tilde{S} u_{Rq} + Y^{pq}_d~\bar{Q_{Lp}} S d_{Rq}  + \text{h.c.} \Big].
\eea
In compliance with MFV, we take $Y_{u}^{pq} = \eta_U \frac{\sqrt{2}m_{u}}{v} \delta^{pq}$ and $Y_{d}^{pq} = \eta_D \frac{\sqrt{2}m_{d}}{v} \delta^{pq}$. We refer to \cite{Manohar:2006ga} for further details. The scaling constants $\eta_U$ and $\eta_D$ are complex in general. However, they are taken real in this study for simplicity.

\section{Constraints applied}\label{constraints}
The 2HDM plus color octet setup is subject to various restrictions from theory and experiments. We discuss them below.

\subsection{Theoretical constraints}
A perturbative theory demands that the magnitudes of the scalar quartic couplings must be $\leq 4\pi$. Next, tree-level unitarity demands that  the $2 \to 2$ matrices constructed out of the tree-level scattering amplitudes involving the various scalar states of the model must have eigenvalues whose magnitudes are $\leq 8\pi$. The following unitarity conditions can be derived for the present framework~\cite{Cheng:2016tlc}.
\besub
\bea
&& \left[\frac{3}{2} (\lambda_1 + \lambda_2) \pm \sqrt{\frac{9}{4} (\lambda_1 - \lambda_2)^2 + (2 \lambda_3 + \lambda_4)^2}\right] \leq 8 \pi, \label{uni_a} \\
&& \left[\frac{1}{2} (\lambda_1  + \lambda_2) \pm \sqrt{\frac{1}{4} (\lambda_1 - \lambda_2)^2 + \lambda_4^2} \right]  \leq 8 \pi, \label{uni_b}  \\
&& \left[\frac{1}{2} (\lambda_1  + \lambda_2) \pm \sqrt{\frac{1}{4} (\lambda_1 - \lambda_2)^2 + \lambda_5^2} \right]  \leq 8 \pi, \label{uni_c}  \\
&& (\lambda_3 + 2 \lambda_4 - 3 \lambda_5)\leq 8 \pi, \label{uni_d} \\
&& (\lambda_3  -  \lambda_5)\leq 8 \pi, \label{uni_e} \\
&& (\lambda_3 +  \lambda_4 )\leq 8 \pi, \label{uni_f} \\
&& (\lambda_3 + 2 \lambda_4 + 3 \lambda_5)\leq 8 \pi, \label{uni_g}  \\
&& (\lambda_3  +  \lambda_5)\leq 8 \pi, \label{uni_h} \\
&&|\nu_1| \leq 2 \sqrt{2} \pi , ~ |\nu_2| \leq 4 \sqrt{2} \pi, ~|\nu_3| \leq 2 \sqrt{2} \pi,  \\
&&|\omega_1| \leq 2 \sqrt{2} \pi , ~ |\omega_2| \leq 4 \sqrt{2} \pi, ~|\omega_3| \leq 2 \sqrt{2} \pi,  \\
&&|\kappa_1| \leq 2 \pi , ~ |\kappa_2| \leq 4 \pi, 
~|\kappa_3| \leq 4 \pi, \\
&&|17 \mu_3 + 13 \mu_4 + 13 \mu_6| \leq 16 \pi, \label{uni_w} \\
&&|2 \mu_3 + 10 \mu_4 + 7 \mu_6| \leq 32 \pi, \label{uni_x} \\
&&|\nu_4 + \nu_5| \lesssim \frac{32 \pi}{\sqrt{15}}, \label{uni_y} \\
&&|\omega_4 + \omega_5| \lesssim \frac{32 \pi}{\sqrt{15}} \label{uni_z}.
\label{unitarity_cond}
\eea
\eesub
Thus, unitarity restricts the magnitudes of the quartic couplings of the model. Eqs.(\ref{uni_a})-(\ref{uni_h}) correspond to the unitarity limit for a pure two-Higgs doublet scenario
\cite{Ginzburg:2005dt,Kanemura:1993hm,Akeroyd:2000wc,Horejsi:2005da,Grinstein:2015rtl,
Cacchio:2016qyh,Gorczyca:2011he}.
We refer to \cite{He:2013tla,Cheng:2016tlc} for more details. Finally, the conditions ensuring a bounded-from-below scalar potential in this model along different directions in the field space are~\cite{Cheng:2018mkc}:
\besub
\bea
&&\mu = \mu_1 + \mu_2 + \mu_6
 + 2(\mu_3 + \mu_4 + \mu_5) > 0, \label{vsca}\\
&&\mu_1 + \mu_2 + \mu_3 + \mu_4 > 0, \label{vscb}
\\
&& 14(\mu_1 + \mu_2) + 5\mu_6 + 24(\mu_3 + \mu_4) 
- 3|2(\mu_1 + \mu_2) - \mu_6| > 0, \label{vscc}\\
&& 5(\mu_1 + \mu_2 + \mu_6) + 6(2\mu_3 + \mu_4 + \mu_5) 
- |\mu_1 + \mu_2 + \mu_6| > 0, \label{vscd}\\
&& \lambda_1 \geq 0 ,~ \lambda_2 \geq 0 ,~ \lambda_3 \geq - \sqrt{\lambda_1 \lambda_2}, \label{vsc1} \\
&& \lambda_3 + \lambda_4 - |\lambda_5| 
\geq - \sqrt{\lambda_1 \lambda_2}, \label{vsc2} \\
&& \nu_1 \geq  -2 \sqrt{\lambda_1 \mu},  \label{vsc3}\\
&&\omega_1 \geq  -2 \sqrt{\lambda_2 \mu}, \label{vsc4} \\
&&\nu_1 + \nu_2 - 2 |\nu_3|  \geq  -2 \sqrt{\lambda_1 \mu}, \label{vsc6}\\
&&\omega_1 + \omega_2 - 2 |\omega_3|  
\geq  -2 \sqrt{\lambda_2 \mu}, \label{vsc7} \\
&&\l_1 + \frac{\mu}{4} + \nu_1 + \nu_2 + 2\nu_3
- \frac{1}{\sqrt{3}}|\nu_4 + \nu_5| > 0, \label{vsc8} \\
&&\l_2 + \frac{\mu}{4} + \omega_1 + \omega_2 + 2\omega_3
- \frac{1}{\sqrt{3}}|\omega_4 + \omega_5| > 0. \label{vsc8}
\label{stability_cond}
\eea
\eesub
Among the above, Eqs.(\ref{vsc1}) and (\ref{vsc2}) correspond to the pure 2HDM. 
The rest of the conditions ensure positivity of the scalar potential in a hyperspace spanned by both colorless as well as colored fields.

\subsection{Higgs signal strengths}

The model also faces restrictions from signal strength measurements in  different decay modes of the 125 GeV Higgs. Denoting the signal strength for the 
channel $p p \to h, ~h \to i$ by $\mu_i$, it is defined as, 
\bea
\mu_i = \frac{\sigma^{\rm{theory}}(pp \rightarrow h)~ {\rm BR^{theory}}(h \rightarrow i)}{\sigma^{\rm{exp}}(pp \rightarrow h)~ {\rm BR^{exp}}(h \rightarrow i)}.
\label{sig-str-1}
\eea
We take $g g \to h$ as the production process at the partonic level. The cross section for the same can be expressed as 
\bea
\sigma(gg \rightarrow h) = \frac{\pi^2}{8 M_h} \Gamma (h \rightarrow gg)~ \delta(\hat{s} - M_h^2)
\label{xsec:gg-h},
\eea
$\sqrt{\hat{s}}$ being partonic centre-of-mass energy. Further, expressing the branching fractions in terms of the decay widths, one rewrites Eq.(\ref{sig-str-1}) as 
\bea 
\mu_i &=& \frac{\Gamma^{\rm{BSM}}_{h \rightarrow gg}}{\Gamma^{\rm{SM}}_{h \rightarrow gg}} ~\frac{\Gamma_i^{\rm{BSM}}}{\Gamma_{\rm{tot}}^{\rm{BSM}}} ~\frac{\Gamma_{\rm{tot}}^{\rm{SM}}}{\Gamma_i^{\rm{SM}}}.
\label{sig-str-3}
\eea
The {\em alignment limit} {\em i.e.} $\alpha = \beta - \frac{\pi}{2}$ is strictly imposed throughout the analysis in which the $h \to WW,ZZ,\tau^+\tau^-$ decay widths at the leading order are identical to the corresponding SM values. Therefore, the signal strength in these channels deviates from the corresponding SM predictions on account of only the additional contribution to the $g g \to h$ amplitude coming from the colored scalars. This is not the case with the $h \to g g, \gamma\gamma$ signal strengths where additional contributions are encountered from the scalar sector. We refer to \cite{Cheng:2016tlc,Cheng:2017tbn,Chakrabarty:2020msl} for relevant formulae on the decay widths for this framework.

The latest data on Higgs signal strengths for $g g \to h$ is summarised in Table \ref{ss}. We combine the data using $\frac{1}{\sigma^2} = 
\frac{1}{\sigma^2_{\text{ATLAS}}} + \frac{1}{\sigma^2_{\text{CMS}}}$ and
$\frac{\mu}{\sigma^2} = 
\frac{\mu_{\text{ATLAS}}}{\sigma^2_{\text{ATLAS}}} + \frac{\mu_{\text{CMS}}}{\sigma^2_{\text{CMS}}}$. The resulting data is used at 2$\sigma$ in our analysis.

\begin{table}[htpb!]
\centering
\begin{tabular}{|c c c|}
\hline
$\mu_i$  & ATLAS & CMS \\ \hline
$ZZ$ & $1.20^{+0.16}_{-0.15}$\cite{ATLAS:2018bsg} & $0.94^{+0.07}_{-0.07}(\text{stat.})^{+0.08}_{-0.07}(\text{syst.})$\cite{CMS:2019chr}\\ \hline
$W^+ W^-$ & $2.5^{+0.9}_{-0.8}$ \cite{Aad:2019lpq} & $1.28^{+0.18}_{-0.17}$\cite{Sirunyan:2018egh}\\ \hline
$\g\g$ & $0.99 \pm 0.14$\cite{Aaboud:2018xdt} & $1.18^{+0.17}_{-0.14}$\cite{Sirunyan:2018ouh}\\ \hline 
$\tau\bar{\tau}$ & $1.09^{+0.18}_{-0.17}(\text{stat.})^{+0.27}_{-0.22}(\text{syst})^{+0.16}_{-0.11}
(\text{theo syst})^{}_{}$\cite{ATLAS:2018lur} & $1.09^{+0.27}_{-0.26}$\cite{Sirunyan:2017khh}\\ \hline
$b\bar{b}$ & $2.5^{+1.4}_{-1.3}$\cite{Aaboud:2018gay} & $1.3^{+1.2}_{-1.1}$\cite{CMS:2016mmc}\\ \hline
\end{tabular}
\caption{Latest limits on the $h$-signal strengths}
\label{ss}
\end{table}

\subsection{Direct search}

Searches for an $H^+$ in the $e^+ e^- \longrightarrow H^+ H^-$ channel at LEP~\cite{Abbiendi:2013hk} has led to the $M_{H^+} > 100$ GeV for all Types of a 2HDM. As for the Type-X, various exclusion limits are rather weak (compared to Type-II, for instance) owing to the suppressed Yukawa couplings of $H,A,H^+$ with the quarks~\cite{Chowdhury:2017aav}. We take $M_H$ = 150 GeV and $M_{H^+} \geq M_H$ to comply with the exclusion constraints. In foresight, we shall also adhere to $M_A > M_h/2$ to evade the limit on BR($h \to A  A$) that can be derived from BR($h_{125} \to AA \to 4\tau, 2\tau 2\mu$)~\cite{CMS:2018qvj}.

We now discuss exclusion constraints on the color octet mass scale. Color-octet resonances have been searched for at the LHC  in the 
$pp \to S \to j j$ \cite{Aad:2014aqa,ATLAS:2016lvi,Khachatryan:2016ecr,Khachatryan:2015dcf} and $pp \to S \to t \bar{t}$ \cite{Aad:2015fna,CMS:2016zte,CMS:2016ehh} channels. Reference \cite{Miralles:2019uzg} recasted the search of colored scalars at the LHC for the Manohar-Wise scenario. The lightest colored scalar was taken to be $S_R$ therein. Since the colored scalars have Yukawa interactions with the quarks, exclusion limits on the color octet mass scale can depend on the strength of such couplings. Reference \cite{Miralles:2019uzg} reported that no clear constraints were derived from the $p p \to S_R \to t \bar{t}$ channel. As for $p p \to S_R t \bar{t} \to t \bar{t} t \bar{t}$, a bound $M_R \gtrsim$ 1 TeV can be derived for $\eta_U \sim \mathcal{O}(1)$. This bound is therefore expected to relax upon lowering $\eta_U$. Another channel is $p p \to S^+ t \bar{b} \to t \bar{b} t \bar{b}$
that leads to a bound of 800 GeV irrespective of the value of $\eta_U$ and $\eta_D \neq 0$. These bounds should apply to $S_I$, the lightest scalar assumed in our case. We take $\eta_U \ll \eta_D$ = 1 in our study in which case maintaining $M_{S_I} \geq$ 800 GeV complies with the direct search constraints.


\subsection{Lepton flavour universality}

Enhanced Yukawa couplings of the $\tau$-lepton potentially modify the $\tau \to \ell \nu \bar{\nu}$ decay rate by virtue of additional contributions stemming from the 2HDM scalars at both tree and loop-levels. This is particularly seen in the lepton-specific case for high $\tan\beta$. We refer to \cite{Chun:2016hzs} for details where this has been studied extensively. Following \cite{Chun:2016hzs}, we have therefore restricted $\tan\beta < 60$ throughout the analysis to comply with lepton flavour universality.

\section{The CDF II and muon $g-2$ excesses}\label{anomalies}

This section discusses how the measured values of the $W$-mass and muon anomalous magnetic moment can be realised in the 2HDM + color octet setup. 
The $W$-mass predicted by a new physics framework can be expressed in terms of its contributions to the oblique parameters $\Delta S$, $\Delta T$ and $\Delta U$ as~\cite{Maksymyk:1993zm}
\bea
M^2_W &=& M^2_{W,\text{SM}} \bigg[1 + \frac{\alpha_{em}}{c^2_W - s^2_W} 
\bigg( -\frac{\Delta S}{2} + c^2_W \Delta T
+ \frac{c^2_W - s^2_W}{4 s^2_W} \Delta U \bigg)  \bigg]
\eea
where $M_{W,\text{SM}}$ is the mass in absence of quantum corrections, and, $c_W$ and $\a_{em}$ respectively denote the cosine of the Weinberg angle and the fine-structure constant. We list below the contributions from the colorless and colored sectors to the $T$-parameter \cite{Peskin:1991sw,Grimus:2008nb} in the alignment limit.
\besub
\bea
\Delta T_{\text{2HDM}} &=& \frac{1}{16 \pi s^2_W M^2_W}\Big[F(M^2_{H^+},M^2_{H}) + F(M^2_{H^+},M^2_{A}) 
 - F(M^2_{H},M^2_{A})\Big] \,, \label{T2HDM} \nonumber \\
\Delta T_S &=& \frac{N_S}{16 \pi s^2_W M^2_W}\Big[F(M^2_{S^+},M^2_{S_R}) + F(M^2_{S^+},M^2_{S_I}) 
 - F(M^2_{S_R},M^2_{S_I})\Big] \label{TS} \,,
\eea
\eesub
where,
\bea
F(x,y) &=&  \frac{x+y}{2} - \frac{xy}{x-y}~{\rm ln} \bigg(\frac{x}{y}\bigg)~~~ {\rm for} ~~~x \neq y \,, \nonumber \\
&=& 0~~~ {\rm for} ~~~ x = y.
\eea
Similarly, the corresponding contributions to the $S$-parameter read
\besub
\bea
\Delta S_{\text{2HDM}} &=& \frac{1}{2\pi} \Big[\frac{1}{6}\text{log}\Big(\frac{M^2_H}{M^2_{H^+}}\Big) - \frac{5}{108} \frac{M^2_H M^2_A}{(M^2_A - M^2_H)^2} \nonumber \\
&&
 + \frac{1}{6}\frac{M^4_A(M^2_A - 3 M^2_H)}{(M^2_A - M^2_H)^3}\text{log}\Big(\frac{M^2_A}{M^2_{H}}\Big) \Big], \\
\Delta S_S &=& \frac{N_S}{2\pi} \Big[\frac{1}{6}\text{log}\Big(\frac{M^2_{S_R}}{M^2_{S^+}}\Big) - \frac{5}{108} \frac{M^2_{S_R} M^2_{S_I}}{(M^2_{S_I} - M^2_{S_R})^2} \nonumber \\
&&
 + \frac{1}{6}\frac{M^4_{S_I}(M^2_{S_I} - 3 M^2_{S_R})}{(M^2_{S_I} - M^2_{S_R})^3}\text{log}\Big(\frac{M^2_{S_I}}{M^2_{S_R}}\Big) \Big].
\eea
\eesub
The total oblique parameter in the present setup is given by the sum of the colorless and colored components, i.e., $\Delta S = \Delta S_{\text{2HDM}} + \Delta S_S$ and $\Delta T = \Delta T_{\text{2HDM}} + \Delta T_S$. The $M_W$ value reported by CDF II can be accommodated by the following ranges~\cite{Asadi:2022xiy,Lu:2022bgw} of $\Delta S$ and $\Delta T$ for $\Delta U=0$:
\bea
\Delta S = 0.15 \pm 0.08,~~\Delta S = 0.27 \pm 0.06,~~\rho_{ST} = 0.93. \label{oblique}
\eea

\begin{figure}
\centering 
\includegraphics[height = 8 cm, width = 8 cm]{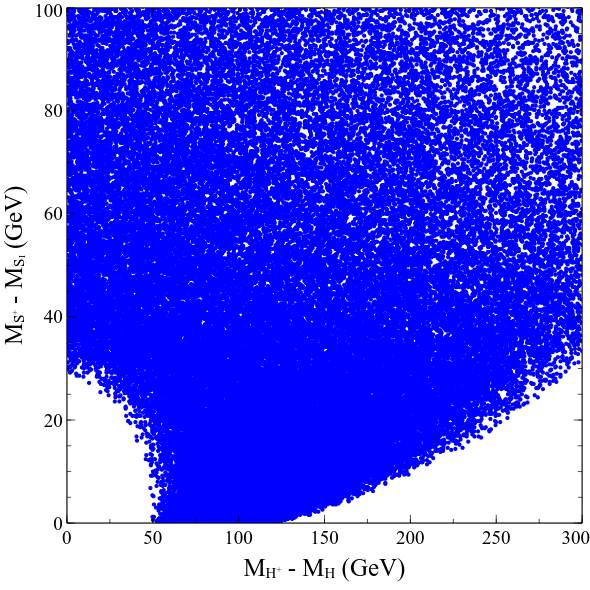}
\caption{Parameter points in the $M_{S^+} - M_{S_I}$ versus $M_{H^+} - M_H$ plane compatible with the observed $M_W$ and the various constraints.}
\label{f:mass_splitting}
\end{figure}

In the above, $\rho_{ST}$ denotes the correlation coefficient. The impact of stipulated ranges for the oblique parameters is expected to get reflected in the scalar mass splittings. To test it, we fix $M_H$ = 150 GeV and $M_{S_I}$ = 800 GeV and make the variations 0 $ < M_{H^+} - M_H < $ 300 GeV, $\frac{M_h}{2} < M_A <$ 200 GeV, 0 $ < M_{S^+} - M_{S_I} < $ 100 GeV and 0 $ < M_{S_R} - M_{S_I} < $ 100 GeV. We plot the parameter points predicting $\Delta S$ and $\Delta T$ in the aforesaid ranges in the $M_{H^+} - M_H$ vs $M_{S^+} - M_{S_I}$ plane in Fig.\ref{f:mass_splitting}. An inspection of the figure immediately suggests that the point $(M_H - M_{H^+},M_{S_I} - M_{S^+})=(0,0)$ is excluded by the CDF data. This is expected on account of the fact that $M_H = M_{H^+}$ and $M_{S_I} = M_{S^+}$ respectively lead to $\Delta T_{\text{2HDM}}$ = 0 and $\Delta T_{S}$ = 0 for all $M_{A}$ and $M_{S_R}$ and a vanishing $\Delta T$ does not suffice to predict the observed $M_W$.

\begin{figure}
\centering 
\includegraphics[height = 8 cm, width = 8 cm]{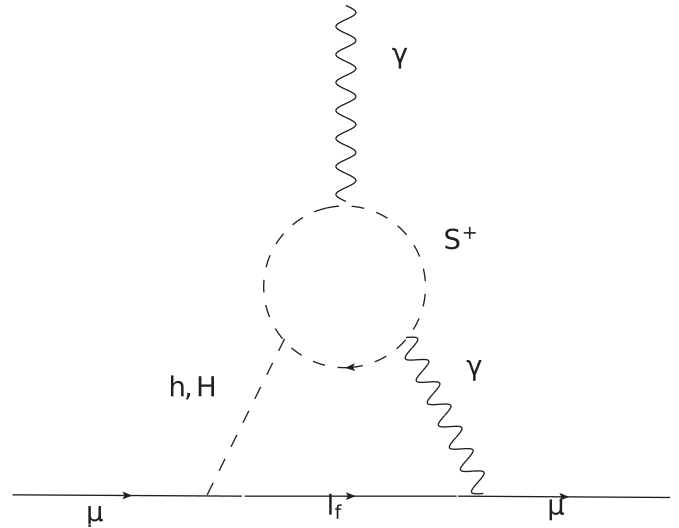}~~~
\includegraphics[height = 8 cm, width = 8 cm]{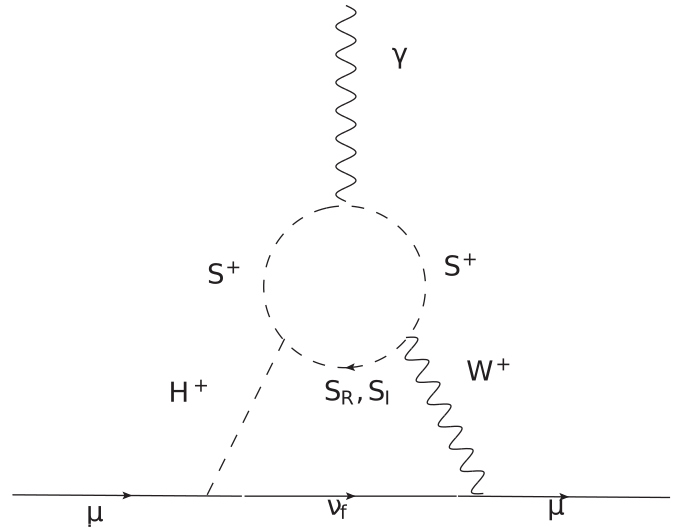}~~~
\caption{Two loop BZ contributions to $\Delta a_\mu$ involving the color octet.}
\label{f:BZ_col_oct_diag}
\end{figure}

We now discuss muon $g-2$ in the given setup. Elaborate discussions on the Type-X 2HDM contributions to $\Delta a_\mu$ are skipped here for brevity. We focus on the contribution coming from the colored scalars in this section.
Since the color-octet does not couple to the leptons at the tree-level, it does not contribute to muon $g-2$ at one-loop.
The color-octet sector contributes to the muon anomalous magnetic moment through the two-loop BZ amplitudes shown in Fig.\ref{f:BZ_col_oct_diag}. The diagram on the left panel is a two-loop topology involving an effective $\phi \gamma \gamma$ ($\phi=h,H$) vertex that is generated at one loop via $S^\pm$ running in the loop. The BZ amplitude can be expressed as
\bea
{\Delta a_\mu}_{\{S^+,~\phi\gamma\gamma\}}^{\text{BZ}} &=& \sum_{\phi = h,H} \frac{N_S\alpha M_\mu^2}{8 \pi^3 M_{\phi}^2}~ y_l^{\phi}~ \lambda_{\phi S^+ S^-}\mathcal{F}\left(\frac{M_{S^+}^2}{M_{\phi}^2}\right). \label{bz1}
\eea
Similarly, the right panel diagram involves an $H^+ W^- \gamma$ vertex that is generated at one loop. The amplitudes stemming from $S_R$ and $S_I$ in the loops are given by
\besub
\bea
{\Delta a_\mu}_{\{S_R,~H^+ W^-\gamma\}}^{\text{BZ}} &=& \frac{N_S \alpha M_\mu^2 }{64 \pi^3 s_w^2 (M_{H^+}^2 - M_W^2)} \zeta_l  ~ \lambda_{ H^+ S^- S_R} \int_{0}^{1} dx~x^2 (x-1) \nonumber \\
&&\times \left[\mathcal{G}\left(\frac{M_{S^+}^2}{M_{H^+}^2},\frac{M_{S_R}^2}{M_{H^+}^2}\right) - \mathcal{G}\left(\frac{M_{S^+}^2}{M_W^2},\frac{M_{S_R}^2}{M_W^2}\right)\right], \label{bz2} \\
{\Delta a_\mu}_{\{S_I,~H^+ W^-\gamma\}}^{\text{BZ}} &=& \frac{N_S\alpha M_\mu^2 }{64 \pi^3 s_w^2 (M_{H^+}^2 - M_W^2)} \zeta_l  ~ \lambda_{ H^+ S^- S_I} \int_{0}^{1} dx~x^2 (x-1) \nonumber \\
&&\times \left[\mathcal{G}\left(\frac{M_{S^+}^2}{M_{H^+}^2},\frac{M_{S_I}^2}{M_{H^+}^2}\right) - \mathcal{G}\left(\frac{M_{S^+}^2}{M_W^2},\frac{M_{S_I}^2}{M_W^2}\right)\right]. \label{bz3} 
\eea
\eesub
The subscripts in Eqs.(\ref{bz1}), (\ref{bz2}) and (\ref{bz3}) refer to the one-loop effective vertex and the circulating colored scalar. The expressions for the trilinear couplings $\l_{\phi S^+ S^-},\lambda_{ H^+ S^- S_R},\lambda_{ H^+ S^- S_I}$ and the functions $\mathcal{F}(z)$ and $\mathcal{G}(z^a,z^b,x)$ are given in the Appendix. We intend to test the magnitudes of the three Barr-Zee contributions and choose tan$\beta$ = 50, $M_{H}$ = 100 GeV, $M_{H^+}$ = 250 GeV, $M_{S_I}$ = 800 GeV, $M_{S^+}$ = 805 GeV, 810 GeV, 820 GeV.
The values taken for tan$\beta$ and $M_{S_I}$ are allowed by the lepton flavour universality and direct search constraints respectively. In addition, the $M_{H^+}-M_H$ and $M_{S^+}-M_{S_I}$ mass differences are thus compatible with $M_W^\text{CDF}$, as can be checked with Fig.\ref{f:mass_splitting}. As for the values of the trilinear couplings, one derives for $\a = \b - \frac{\pi}{2}$ that $\l_{H S^+ S^-} = -\frac{1}{2}\big((\nu_1 - \omega_1)c_\beta s_\beta + \kappa_1 s_{2\beta}\big) \simeq -\frac{\kappa_1}{2}$ for large tan$\beta$. Since $\kappa_1$ is a priori a free parameter of the theory, $|\l_{H S^+ S^-}|$ can be as large as 2$\pi$. It similarly follows that $|\l_{H^+ S^- S_R}|$ and $|\l_{H^+ S^- S_I}|$ $\lesssim \pi$. 

\begin{figure}[htpb!]
\centering 
\includegraphics[height = 8 cm, width = 8 cm]{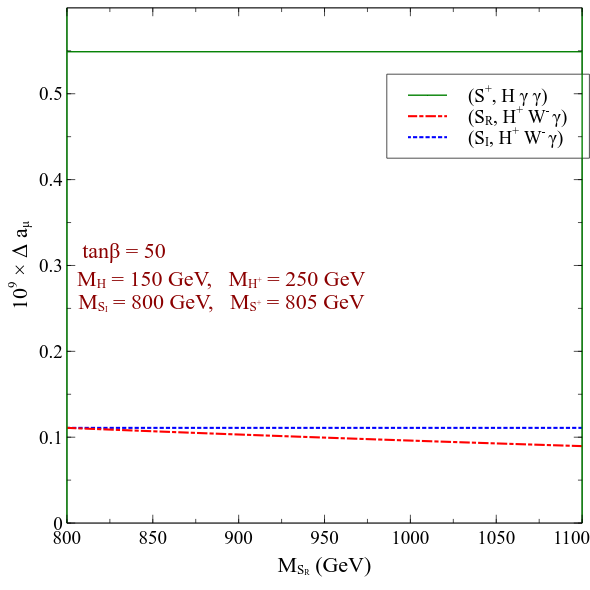}~~~
\includegraphics[height = 8 cm, width = 8 cm]{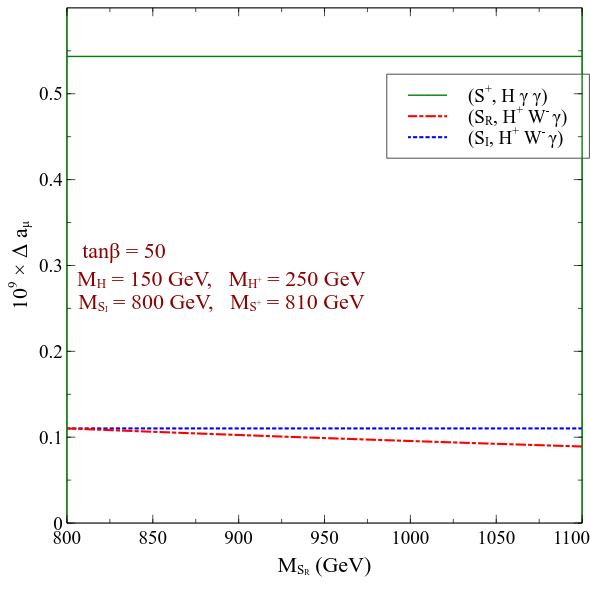}\\
\includegraphics[height = 8 cm, width = 8 cm]{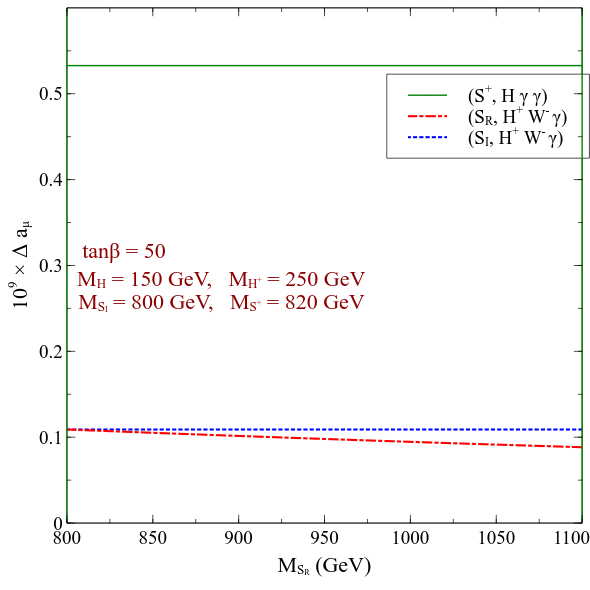}
\caption{Variation of different BZ contributions involving colored scalars for $M_{S^+}$ = 805 GeV (top left), 810 GeV (top right) and 820 GeV (bottom).}
\label{f:BZ_col_oct}
\end{figure}

We plot the individual BZ amplitudes in Fig.\ref{f:BZ_col_oct} versus $M_{S_R}$ tan$\beta$ = 50 and $\l_{H S^+ S^-} = -2\pi$ and $\l_{H^+ S^- S_R}=\l_{H^+ S^- S_I} = -\pi$. We find that they can be $\mathcal{O}(10^{-10})$ with the largest being
${\Delta a_\mu}_{\{S_R,~H^+ W^-\gamma\}}^{\text{BZ}}$. This sizeable magnitudes can be understood from the fact that the products $\l_{H S^+ S^-}\times \tan\beta$, $\l_{H^+ S^- S_R}\times\tan\beta$ and $\l_{H^+ S^- S_I}\times\tan\beta$ are $\mathcal{O}(100)$ numbers. Variations introduced by the said changes of $M_{S^+}$ are small and do not change the ball-park contributions to $\Delta a_\mu$. 

\begin{figure}[htpb!]
\centering 
\includegraphics[height = 7 cm, width = 7 cm]{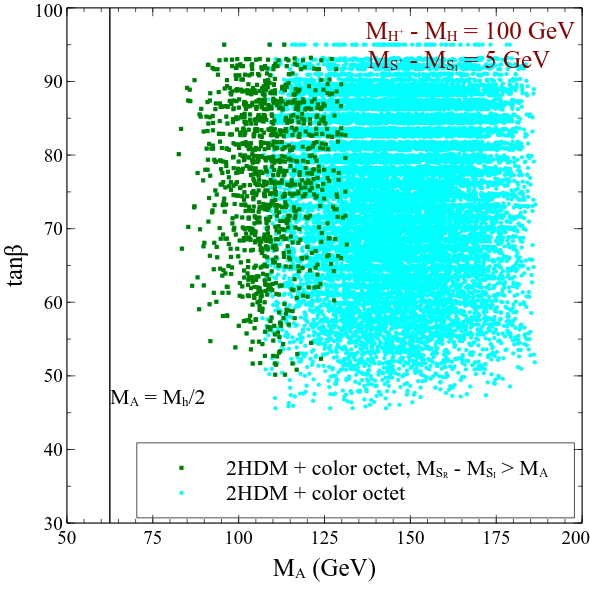}
\includegraphics[height = 7 cm, width = 7 cm]{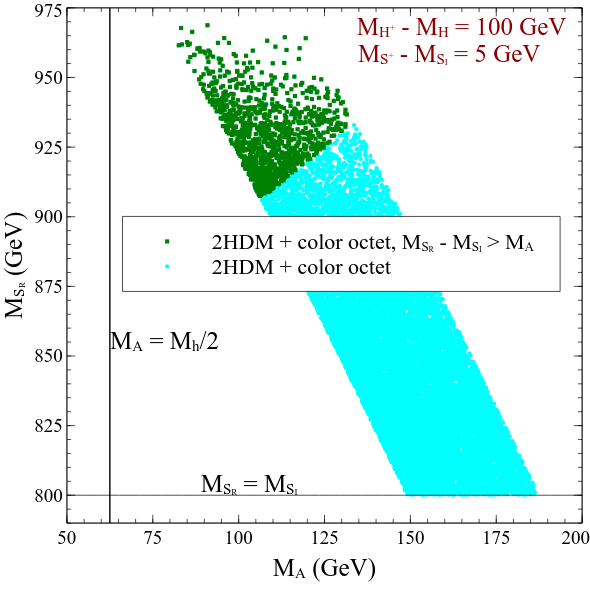} \\
\includegraphics[height = 7 cm, width = 7 cm]{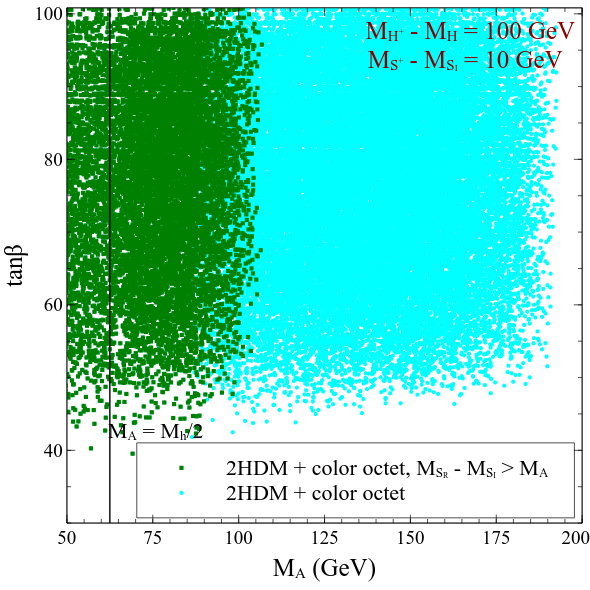}
\includegraphics[height = 7 cm, width = 7 cm]{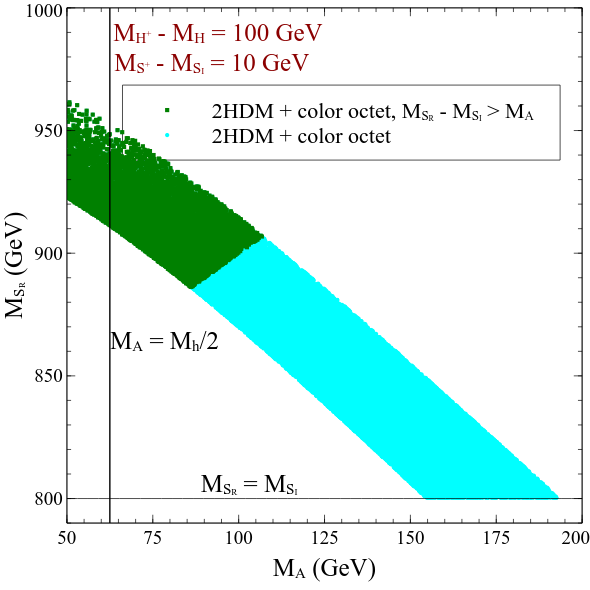} \\
\includegraphics[height = 7 cm, width = 7 cm]{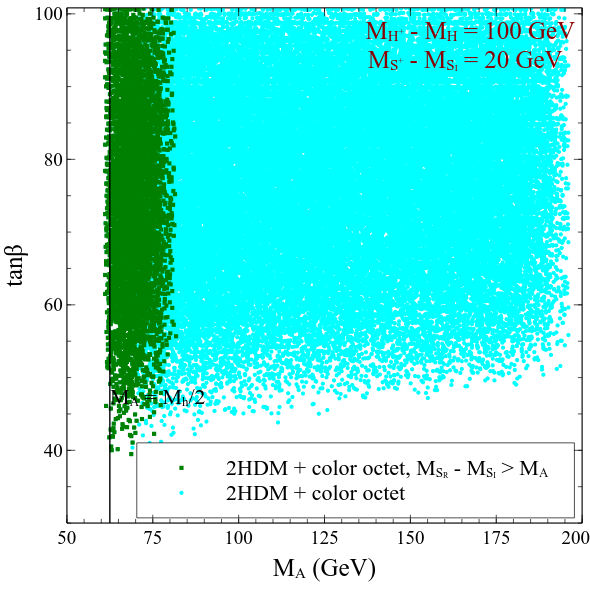}
\includegraphics[height = 7 cm, width = 7 cm]{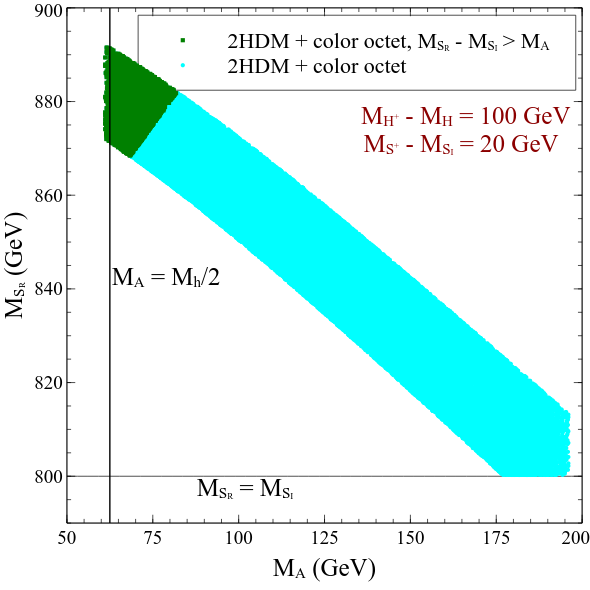} \\
\caption{Variation of ${\Delta a_\mu}_{\{S^+,~\phi\gamma\gamma\}}^{\text{BZ}}$ for $M_{S^+}$ = 805 GeV (top left), 810 GeV (top right) and 820 GeV (bottom).}
\label{f:param_col_oct}
\end{figure}

Retaining the same values for the scalar masses as in Fig.\ref{f:BZ_col_oct}, we perform the following scan over the rest of the parameters:
\bea
20~\text{GeV} < M_A <  200~\text{GeV},~0 < m_{12} < 100~\text{GeV}, \nonumber \\
10 < \tan\beta < 100,~|\omega_1|,|\kappa_1|,|\kappa_2|,|\kappa_3|, |\nu_1|,|\nu_2|,|\nu_3| < 2\pi.
\eea
Parameter points that negotiate all constraints successfully and are consistent with the observed muon $g-2$ and $M_W$ at $2\sigma$ and $3\sigma$ respectively are plotted in the $M_A-\tan\beta$ ($M_A-M_{S_R}$) plane in the left (right) panel of Fig.\ref{f:param_col_oct}. One inspects in this figure that on account of the color-octet contributions, an $A$ that is compatible with $\Delta a_\mu$ can now be much heavier compared to what it is in the pure Type-X 2HDM. To elucidate, the enlarged parameter space now includes $M_A \lesssim 180$ GeV for a tan$\beta$ around 50 for the all three $M_{S^+}$ values taken. The lower bound $M_A \gtrsim 80$ GeV is noticed for $M_{S^+}$ = 805 GeV. This is a consequence of demanding $\Delta T$ and $\Delta S$ in the stated ranges (Eq.(\ref{oblique})) so as to comply with the observed $M_W$. We also show the subregions where $M_{S_R} > M_{S_I} + M_A$ on account of the $S_R \to S_I A$ decay signal in foresight. Such a requirement restricts $M_A \lesssim$ 140 GeV, 110 GeV and 85 GeV for $M_{S^+}$ = 805 GeV, 810 GeV and 820 GeV respectively.                    
 
 \section{Collider Analysis} 
\label{collider}

Having discussed the features of the multi-dimensional parameter space validated through the theoretical and experimental constraints, in this section, we aim to analyse a promising signature involving the non-standard colored scalars in high-luminosity (HL) 14 TeV LHC. The signal topology allows for the single production of $S_R$ dominantly through gluon-gluon and quark fusion  and then subsequent decay of $S_R$ into $S_I$ and $A$. Finally the colored scalar $S_I$ decays into two $b$-jets and $A$ decays to $\tau^+ \tau^-$. In short we shall be analysing  the following signal :
\bea
p p \to S_R \to S_I A , S_I \to b \bar{b}, A \to \tau^+ \tau^- \,.
\label{signal-proc}
\eea
Depending on the visible decay products of the $\tau^+$ and $\tau^-$ in the final state, there could be three different final states :
\begin{itemize}
\item Both $\tau$ leptons in the final state decay leptonically leading to the final state $2 \tau_\ell + 2b + \met$ with $\tau_\ell = \tau_e, \tau_\mu$. In short we shall denote this case as ``DL".
\item One of the two $\tau$s in the final state decays leptonically, whereas the second one decays hadronically. This semi-leptonic decay topology gives rise to $1 \tau_\ell + 1 \tau_h + 2b + \met$ final state. For convenience this case will be denoted as ``SL".
\item Both $\tau$ leptons decaying hadronically {\footnote{The visible decay product of the hadronic decay of $\tau$-lepton is identified as $\tau$-jet.}}, give rise to $2 \tau_h + 2 b + \met$ final state. Since there is no lepton in the final state, this case will be denoted by "NoL".
\end{itemize}

Now to make $S_R \to S_I A$ decay mode mentioned in Eq.(\ref{signal-proc}) kinematically open, we need to set $M_{S_R} > M_{S_I} + M_A$.
We throughout take $\eta_D=1$ and $\eta_U \ll \eta_D$ for the collider analysis. We also fix $M_{S_I} = 800$ GeV which is then compatible with the direct search constraints discussed in section \ref{constraints}. Next, we choose five benchmark points (BP1-BP5) characterized by low, medium and high masses of $A$ ranging from 66 GeV to 147 GeV. All the benchmarks are not only allowed by the theoretical and experimental constraints, but also can envisage the muon anomalous magnetic moment within the $3 \sigma$ band about the central value and address the $W$-mass anomaly simultaneously. For the chosen benchmarks, the masses of other scalars like $H^+, S^+$, the branching ratios of the processes $S_R \to S_I A, ~S_I \to b \bar{b},~ A \to \tau^+ \tau^-$ along with the corresponding values of $\Delta a_\mu$, $(M_W^{\rm CDF}- 80.000)$ are tabulated in Table \ref{bsm}. BR$(S_R \to S_I A)$ is $\sim 99\%$ for BP1 and BP2. Since the mass splitting $(M_{S_R}-M_{S_I})$ increases from BP3 to BP5, accordingly $S_R \to S_I Z ,~ S_R \to S^\pm W^\mp$ decay modes  open up leading to the decrease in BR$(S_R \to S_I A)$. For all benchmarks BR$(A \to \tau^+ \tau^-)$ are almost $\sim 99 \%$. Lastly, our choice of $\eta_{U,D}$ ensures that $S_I \to b \bar{b}$ is the dominant decay mode.

\begin{table}[htpb!]
\begin{center}
\resizebox{16cm}{!}{
\begin{tabular}{ |c | c | c | c | c | c | c | c | c | c | c| c| } 
\hline
 & tan$\beta$ & $M_A$ (GeV)& $M_{H^+}$ (GeV) & $M_{S_R}$ (GeV) & $M_{S_I}$ (GeV) & $M_{S^+}$ (GeV) & BR$(S_R \to S_I A)$ 
 & BR$(S_I \to b \bar{b})$ & BR$(A \to \tau^+ \tau^-)$ & $\Delta a_\mu \times 10^9$ & $(M^{\text{CDF}}_W - 80.000)$ (MeV)  \\ \hline
BP1 & 43.264 & 66.39 & 250.0 & 876.994 & 800.0 & 820.0 & 0.998653 & 0.866694 & 0.996484 & 0.77824 (3$\sigma$) & 433.573 \\ \hline 
BP2 & 56.075 & 80.093 & 250.0 & 882.644 & 800.0 & 820.0 & 0.994456 & 0.866694 & 0.996488 & 0.74883 (3$\sigma$) & 417.401 \\ \hline
BP3 & 55.565 & 100.314 & 250.0 & 909.707 & 800.0 & 810.0 & 0.791145 & 0.866694 & 0.996489 & 0.77966 (3$\sigma$) & 418.839 \\ \hline
BP4 & 54.48 & 121.11 & 250.0 & 938 & 800 & 805 & 0.484672 & 0.866694 & 0.99649 & 0.77224 (3$\sigma$) & 423.641 \\ \hline
BP5 & 58.7 & 147.0 & 250.0 & 950.3 & 800 & 800 & 0.157716 & 0.866694 & 0.996491 & 0.75824 (3$\sigma$) & 444.802 \\ \hline
\end{tabular}}
\caption{Benchmarks compatible with $M^{\text{CDF}}_W$ and the observed $\Delta a_\mu$.}
\label{bsm}
\end{center}
\end{table}

Next we discuss the relevant backgrounds corresponding to the signals mentioned earlier. The dominant contributors to the backgrounds are $p p \to Z \to \tau^+ \tau^- + jets, ~ p p \to t \bar{t} \to 1 \ell + jets,~ ~ p p \to t \bar{t} \to 2 \ell + jets$ \footnote{All the background samples having jets in the final state are generated by matching the samples up to two jets.}. The first background mimics the signal if the light jets are faked as $b$-jets. In the second background, if one of the light jets is mis-tagged as a $\tau$-jet and two of the light jets are faked as $b$-jets, then the final state resembles with $1 \tau_h + 1 \tau_\ell + 2b + \met.$ In addition to the previously mentioned conditions in the second background, one of the leptons should be missed in order to achieve the same signal topology. Apart from these, there are several sub-dominant background processes like $tW,~ WZ \to 2 \ell 2q,~ WZ \to 3 \ell \nu + jets$ etc. A complete set of all possible backgrounds can be found in Table \ref{tab:xsecs}.

The particle interactions relevant for collider analysis are first implemented in \texttt{FeynRules} \cite{Alloul:2013bka}. As an output, the Universal Feynrules Output (UFO) file is generated. The signal and background cross sections are calculated at the leading order (LO) through \texttt{MG5aMC@NLO} \cite{Alwall:2014hca} using the aforesaid UFO file. Further showering and hadronization are done via \texttt{Pythia8} \cite{Sjostrand:2014zea}. To incorporate the detector effect we use the default CMS detector simulation card included in Delphes-3.4.1~\cite{deFavereau:2013fsa}. We have used \texttt{anti-}$k_t$ jet-clustering algorithm \cite{Cacciari:2008gp} for jet reconstruction. We shall analyse the signal using both traditional cut-based method and sophisticated machine learning techniques in this study. We expect an improvement in the result while performing the later. The signal significance $\mathcal{S}$ can be calculated in terms of the number of signal ($S$) and background events ($B$) left after imposing relevant cuts using : $\mathcal{S} = \frac{S}{\sqrt{B}}$. After taking into account $\theta \%$ systematic uncertainty, the significance turns out to be $\mathcal{S} =\frac{S}{\sqrt{B+ (\theta*B/100)^2}}$ \cite{Cowan:2010js}. The signal and background cross sections at LO are computed using \texttt{MG5aMC@NLO}. While evaluating the cross sections, for some of the backgrounds (mentioned in Table \ref{tab:xsecs}), we use the acceptance cuts (tabulated in Table \ref{tab:objSel} and mentioned in item C0 later) at the generation level. For other backgrounds, we impose the similar cuts at the detector level to keep all the event samples at the same footing. The LO cross sections of some of the backgrounds are multiplied with relevant $k$-factors to obtain NLO cross sections. The signal and background cross sections are tabulated in Table \ref{tab:xsecs}.

\begin{table}[htpb!]
  \begin{center}
    {\footnotesize
    \begin{tabular}{|l|c|}
      \hline
      Process                                            &         cross section (pb)         \\ 
      \hline\hline
      \multicolumn{2}{|l|}{\texttt{Signal benchmarks}} \\
      \hline\hline
      \texttt{BP1}                                       &              $0.0431$               \\
      \texttt{BP2}                                       &              $0.0429$               \\
      \texttt{BP3}                                       &              $0.0342$               \\
      \texttt{BP4}                                       &              $0.0209$               \\
      \texttt{BP5}                                       &              $0.0068$               \\
      \hline\hline
      \multicolumn{2}{|l|}{\texttt{SM Backgrounds}} \\
      \hline\hline
      \texttt{$t\bar{t}\,\to\,2\ell\,+\,jets$}           &    $107.65$ [NNLO] \\
      \texttt{$t\bar{t}\,\to\,1\ell\,+\,jets$}           &    $437.14$ [NNLO]     \\
      \texttt{$tW$}                                      &    $34.81$ [LO]                    \\
      \texttt{$Z\,\to\,\tau^+\tau^-\,+\,jets$}           &    $803$ [NLO]                     \\
      \texttt{$t\bar{t}W\,\to\,\ell\nu\,+\,jets$}        &    $0.25$ [NLO]                    \\
      \texttt{$t\bar{t}W\,\to\,qq$}                      &    $0.103$ $^1$ [LO]               \\
      \texttt{$t\bar{t}Z\,\to\,\ell^+\ell^-\,+\,jets$}   &    $0.24$ [NLO] \cite{Kardos:2011na}\\
      \texttt{$t\bar{t}Z\,\to\,qq$}                      &    $0.206$ $^1$ [NLO]  \cite{Kardos:2011na}   \\
      \texttt{$WZ\,\to\,3\ell\nu\,+\,jets$}              &    $2.27$ [NLO]  \cite{Campbell:2011bn}        \\
      \texttt{$WZ\,\to\,2\ell\,2q$}                      &    $4.504$ [NLO] \cite{Campbell:2011bn}         \\
      \texttt{$ZZ\,\to\,4\ell$}                          &    $0.187$ [NLO]    \cite{Campbell:2011bn}     \\
      \texttt{$t\bar{t}h\,\to\,\tau^+\,\tau^-$}          &    $0.006$ $^1$ [LO]                           \\
      \texttt{$b\bar{b}\tau^+\tau^-$}                    &    $0.114$ $^1$ [LO]                           \\
      \texttt{$WWW$}                                     &    $0.236$ [NLO]                               \\
      \texttt{$WWZ$}                                     &    $0.189$ [NLO]                               \\
      \texttt{$WZZ$}                                     &    $0.064$ [NLO]                               \\
      \texttt{$ZZZ$}                                     &    $0.016$ [NLO]                               \\
      \hline
    \end{tabular}
    }
  \end{center}
  \footnotesize{
    $^1$ Some selections are applied at the generation (i.e. Madgraph) level. $p_T$ of jets(j) and $b$ quarks(b) $>\,20$ GeV, $p_T$ of leptons($\ell$) $>\,10$ GeV, $|\eta|_{j/b}\,<\,5$, $|\eta|_\ell\,<\,2.5$ and $\Delta R_{jj/\ell\ell/j\ell/b\ell}\,>\,0.4$.  \\
  }
  \caption{Cross sections of the signal benchmark points and the relevant SM backgrounds.}
  \label{tab:xsecs}
\end{table}

The subsequent discussion is divided into the two following subsections that contain cut-based and multivariate analyses respectively.


\subsection{Cut-based analysis}
We first apply a few pre-selection cuts (C0-C4) on the events which are used as baseline selection criterion and then perform cut based as well as multivariate analyses to estimate the signal sensitivity. Below we describe the baseline selection criterion in detail.

\begin{itemize}
  
\item[C0:] A few basic selection criteria are applied to select $e, \mu, \tau $ and jets in the final state. We construct the following set of kinematic variables both for leptons and jets: $(a)$ transverse momentum $p_T$, $(b)$ pseudo-rapidity $\eta$, and $(c)$ separation between $i$ and $j$-th objects $\Delta R_{ij}\,=\,\sqrt{(\Delta \eta_{ij})^2 + (\Delta \Phi_{ij})^2}$, which is defined in terms of the azimuthal angular separation $(\Delta \Phi_{ij})$ and pseudo-rapidity difference $(\Delta \eta_{ij})$ between the same objects. The chosen threshold values of these variables are quoted in Table~\ref{tab:objSel}.

  \begin{table}[htpb!]
    \begin{center}
      \footnotesize\setlength{\extrarowheight}{2pt}
      \begin{tabular}{|l|l|}
        \hline Objects & Selection cuts \\
        \hline
        \texttt{$e$}          & $p_{T} > 10$~{\rm GeV}, $~|\eta| < 2.5$                                              \\
        \texttt{$\mu$}        & $p_{T} > 10$~{\rm GeV}, $~|\eta| < 2.4$, $~\Delta R_{\mu e} > 0.4$                   \\
        \texttt{$\tau_{h}$}   & $p_{T} > 20$~{\rm GeV}, $~|\eta| < 2.4$, $~\Delta R_{\tau_h e/\mu} > 0.4$             \\
        \texttt{$b\,jets$}    & $p_{T} > 20$~{\rm GeV}, $~|\eta| < 2.5$, $~\Delta R_{{\rm b\,jet}~ e/\mu} > 0.4$     \\
        \hline
      \end{tabular}
      \footnotesize
    \end{center}
    \caption{ Summary of acceptance cuts to select analysis level objects}
    \label{tab:objSel}
  \end{table}
  
\item[C1:] Next we ensure that the final state acquires correct lepton multiplicity. By lepton, here we mean $\mu$ and $e$ only. In the final state, we demand one and zero leptons for SL and NoL channels respectively.
\item[C2:] As expected from the topology of the signals, we require two $\tau$-jets for NoL channel. Similarly, for SL channel, one $\tau$-jet is demanded.
\item[C3:] Since the lepton and the $\tau$-jet (two $\tau$-jets)  originate from two oppositely charged $\tau$-leptons in the SL (NoL) channel, the decay products in both cases must possess opposite charges. We impose this particular condition to the charges of the final particles. 
\item[C4:] Since the signals in both channels include two $b$-jets in the final state coming from $S_R$, we demand two $b$-jets in the final state for both channels. 
  
\end{itemize}

\begin{figure}[htpb!]{\centering
\subfigure[SL]{
\includegraphics[height = 4.5 cm, width = 8.0 cm]{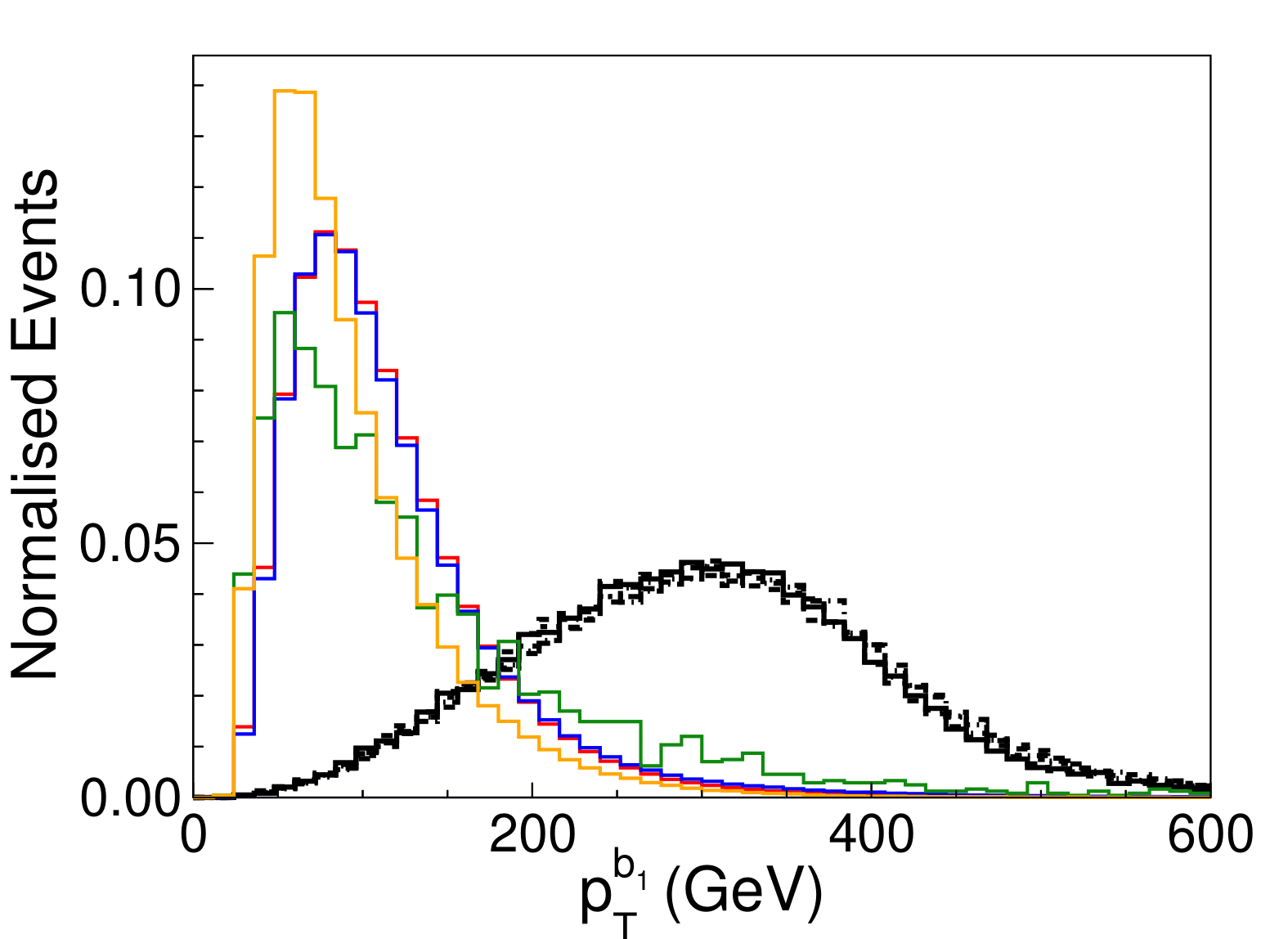}}
\subfigure[NoL]{
\includegraphics[height = 4.5 cm, width = 8.0 cm]{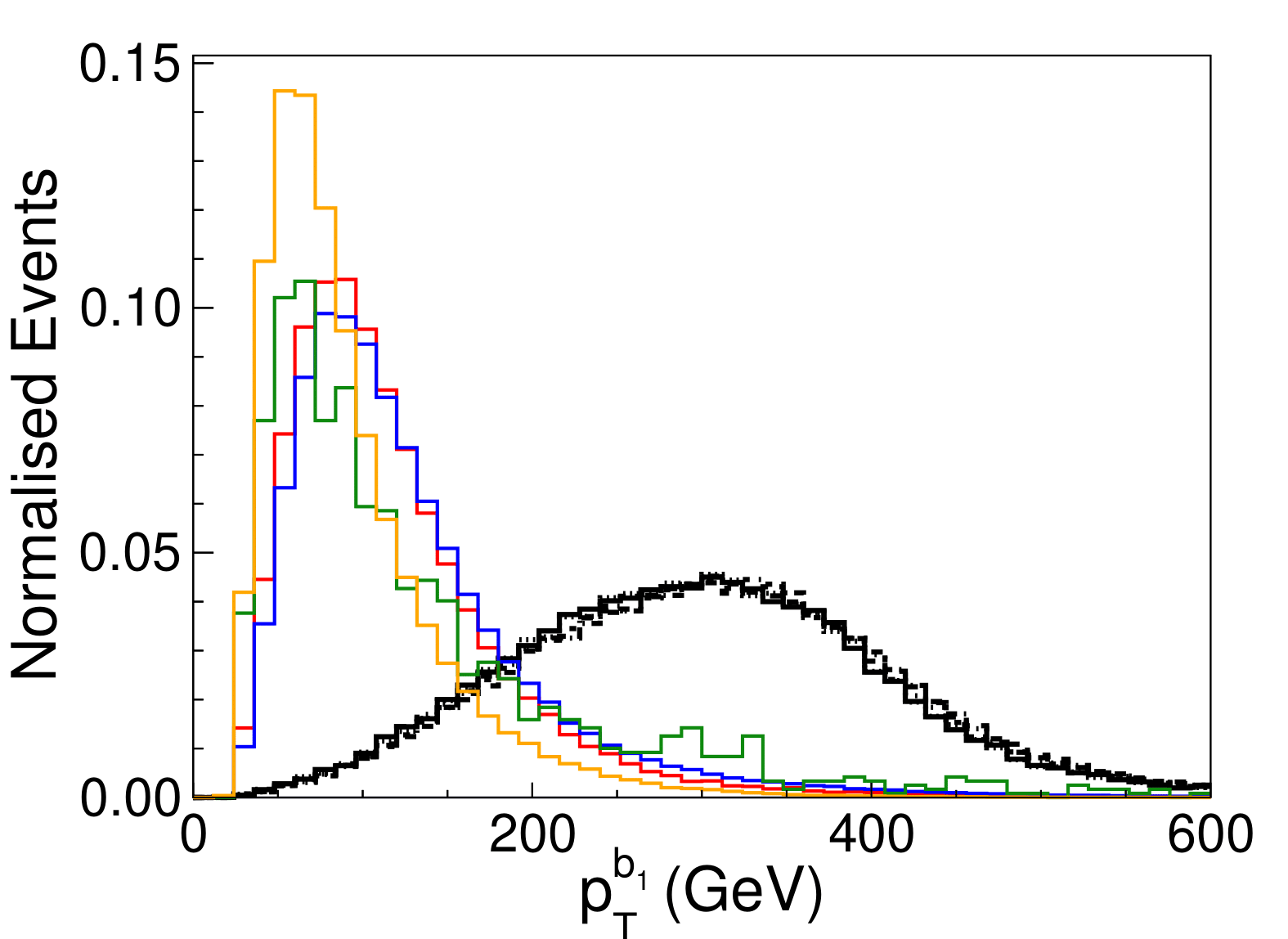}} \\
\subfigure[SL]{
\includegraphics[height = 4.5 cm, width = 8.0 cm]{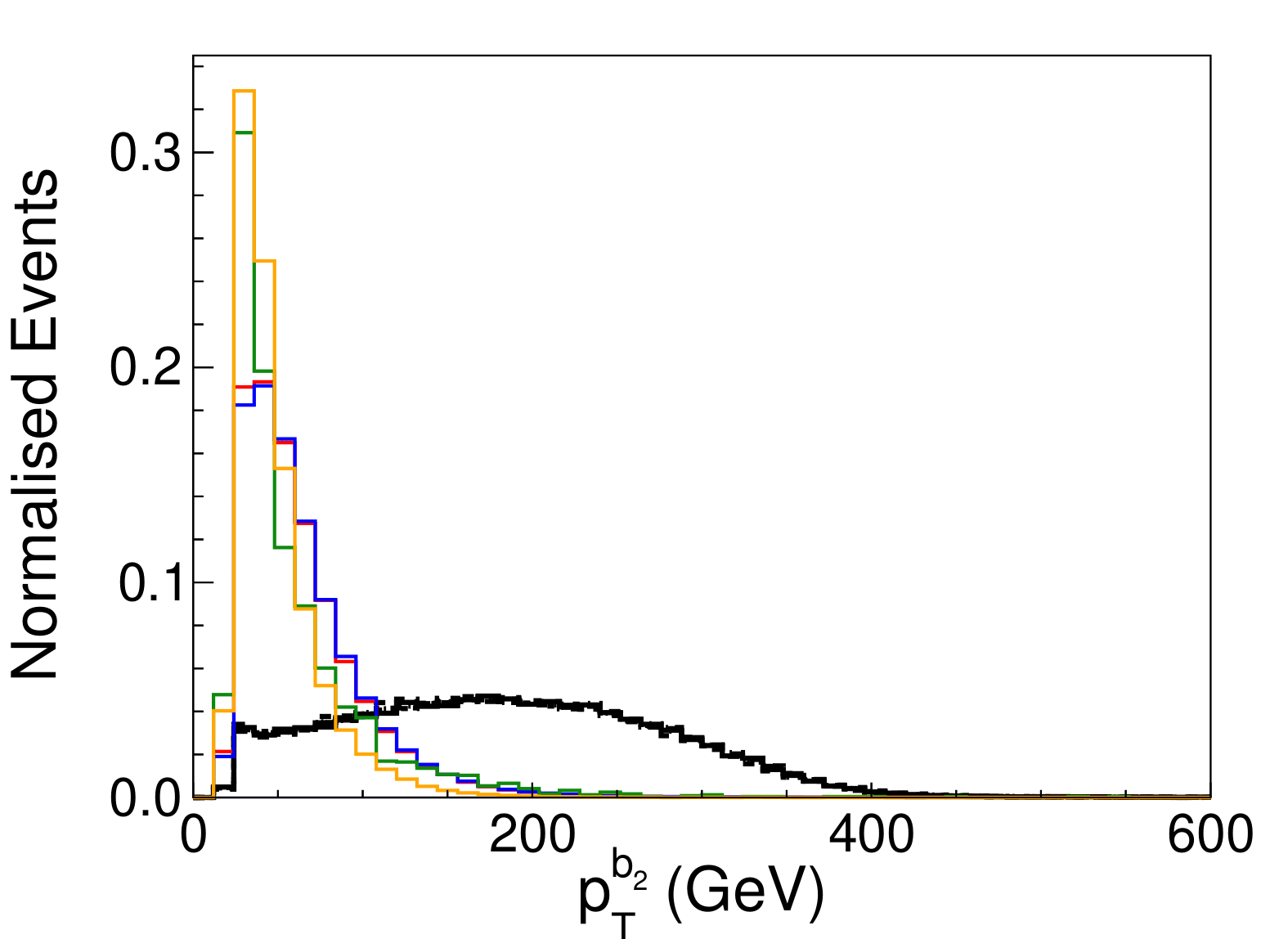}}
\subfigure[NoL]{
\includegraphics[height = 4.5 cm, width = 8.0 cm]{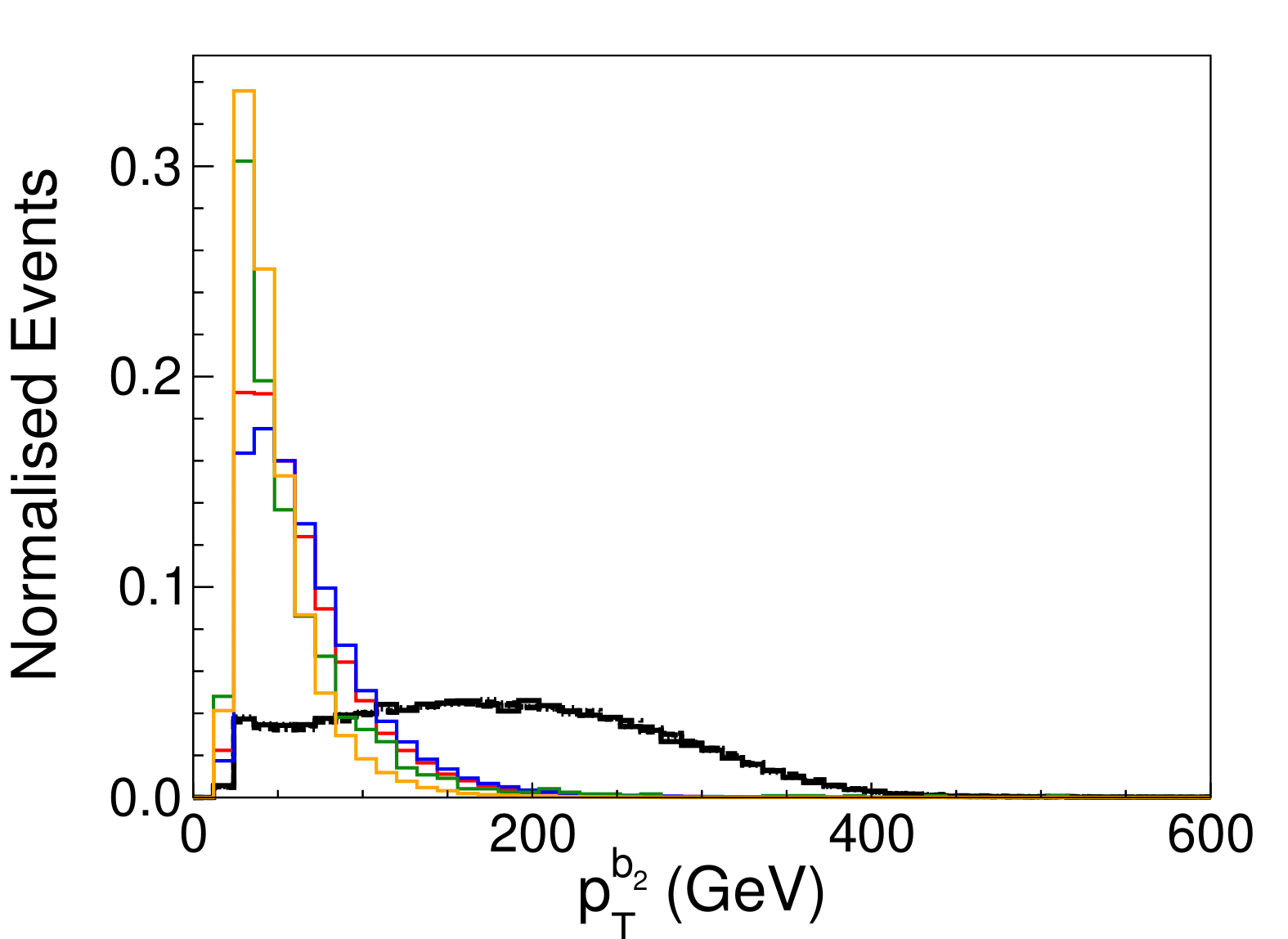}} \\
\subfigure[SL]{
\includegraphics[height = 4.5 cm, width = 8.0 cm]{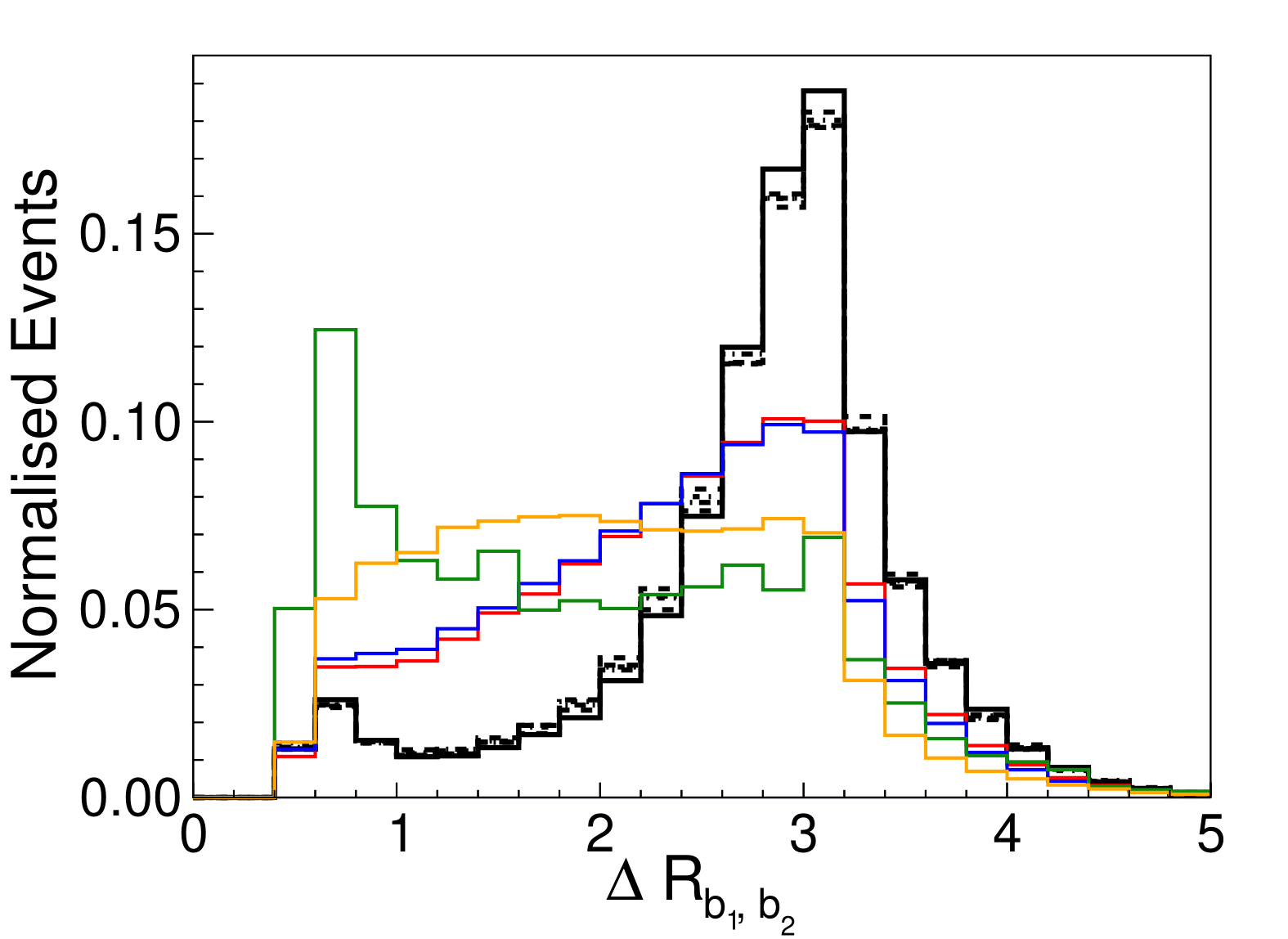}}
\subfigure[NoL]{
\includegraphics[height = 4.5 cm, width = 8.0 cm]{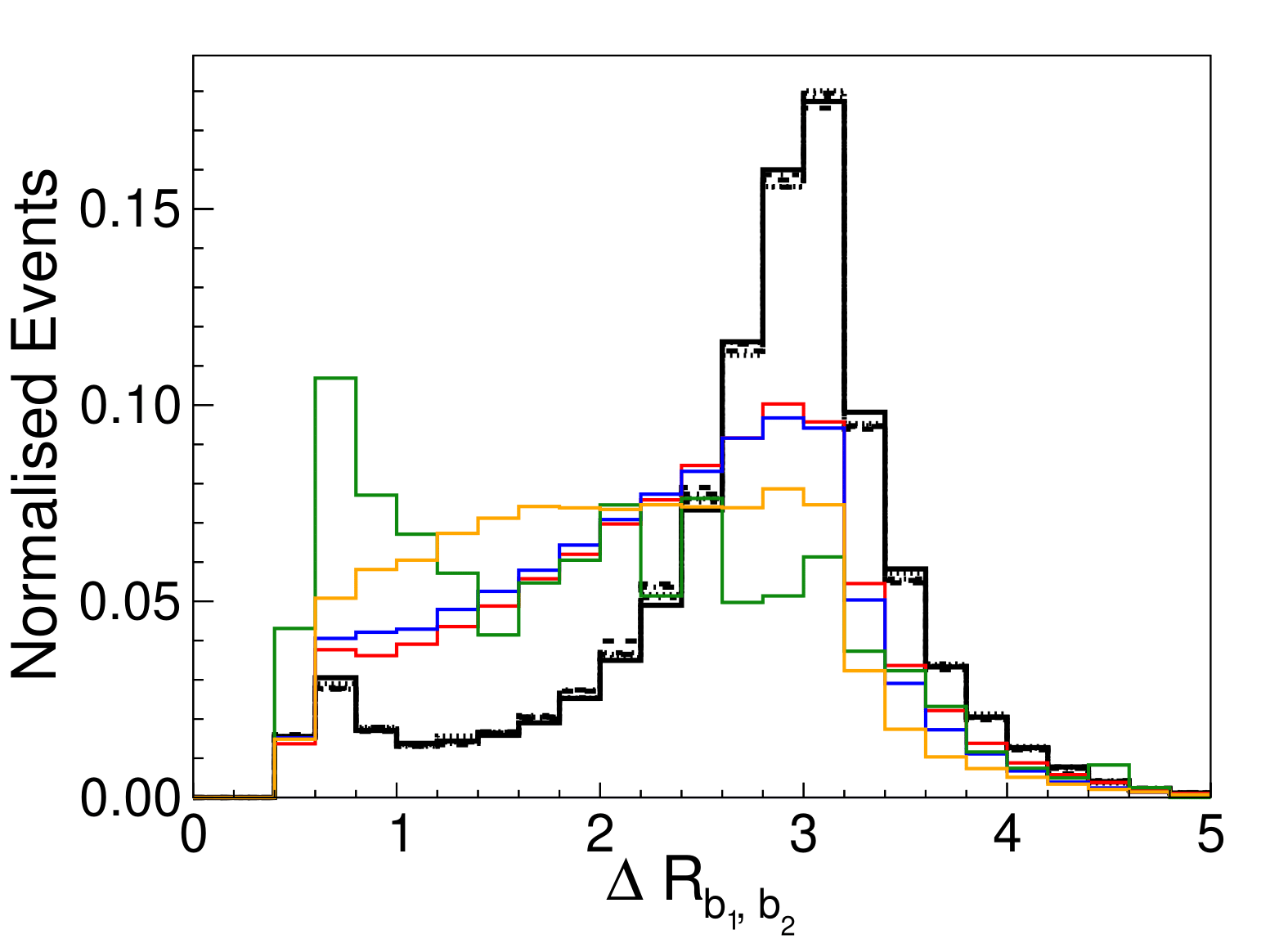}} 
} 
 \hspace{0.01\textwidth}
  \label{fig:leg}
  \centering
  \includegraphics[width=0.7\textwidth]{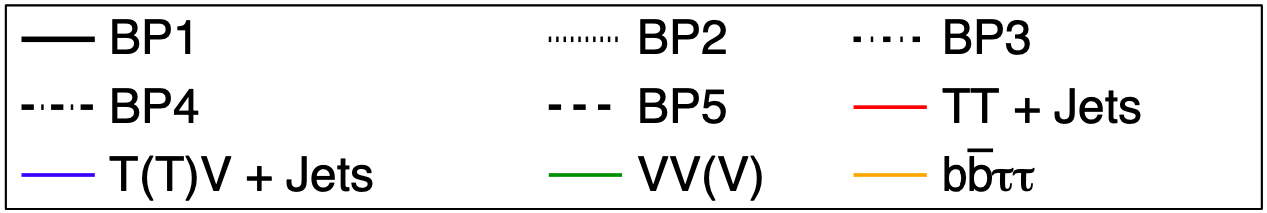}
\caption{ Distributions of some kinematic variables: (a,b) Distribution of leading b jet $p_T$, (c,d) Distributions of sub-leading b jet $p_T$, (e,f) $\Delta R$ between two $b$-jets for SL and NoL channels respectively.}
\label{fig:features-1}
\end{figure}

\begin{figure}[htpb!]{\centering
\subfigure[SL]{
\includegraphics[height = 4.5 cm, width = 8.0 cm]{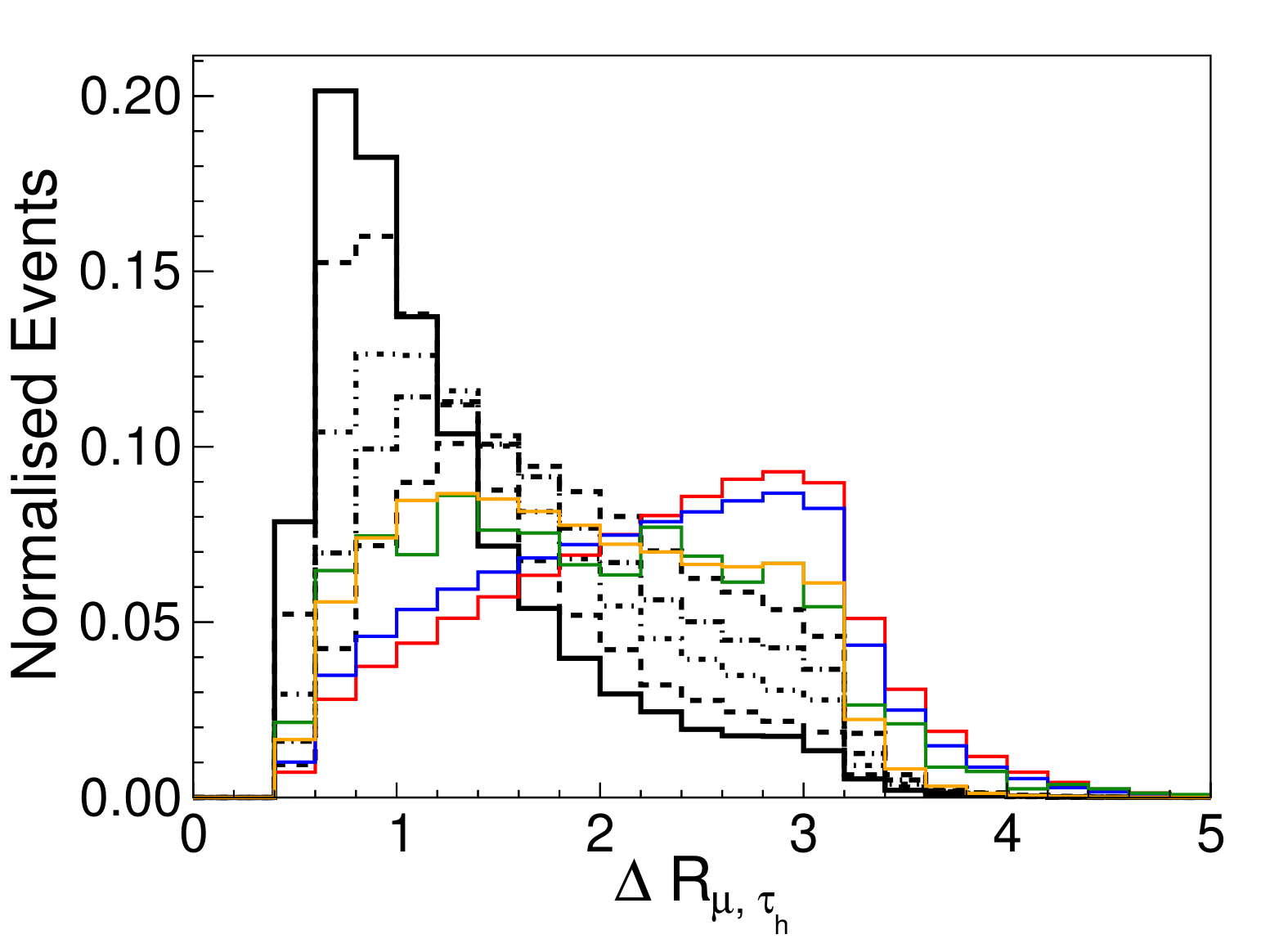}}
\subfigure[NoL]{
\includegraphics[height = 4.5 cm, width = 8.0 cm]{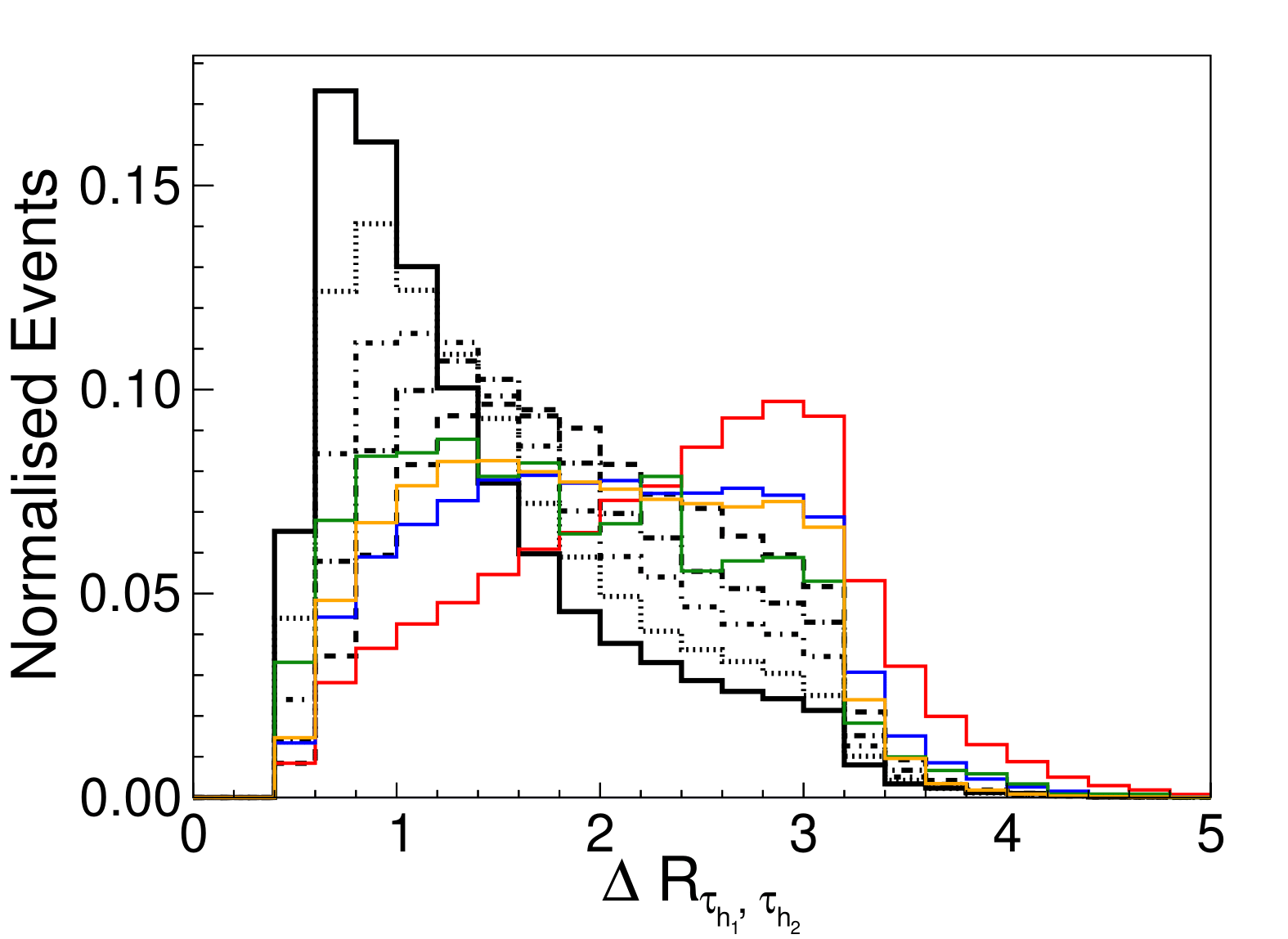}} \\
\subfigure[SL]{
\includegraphics[height = 4.5 cm, width = 8.0 cm]{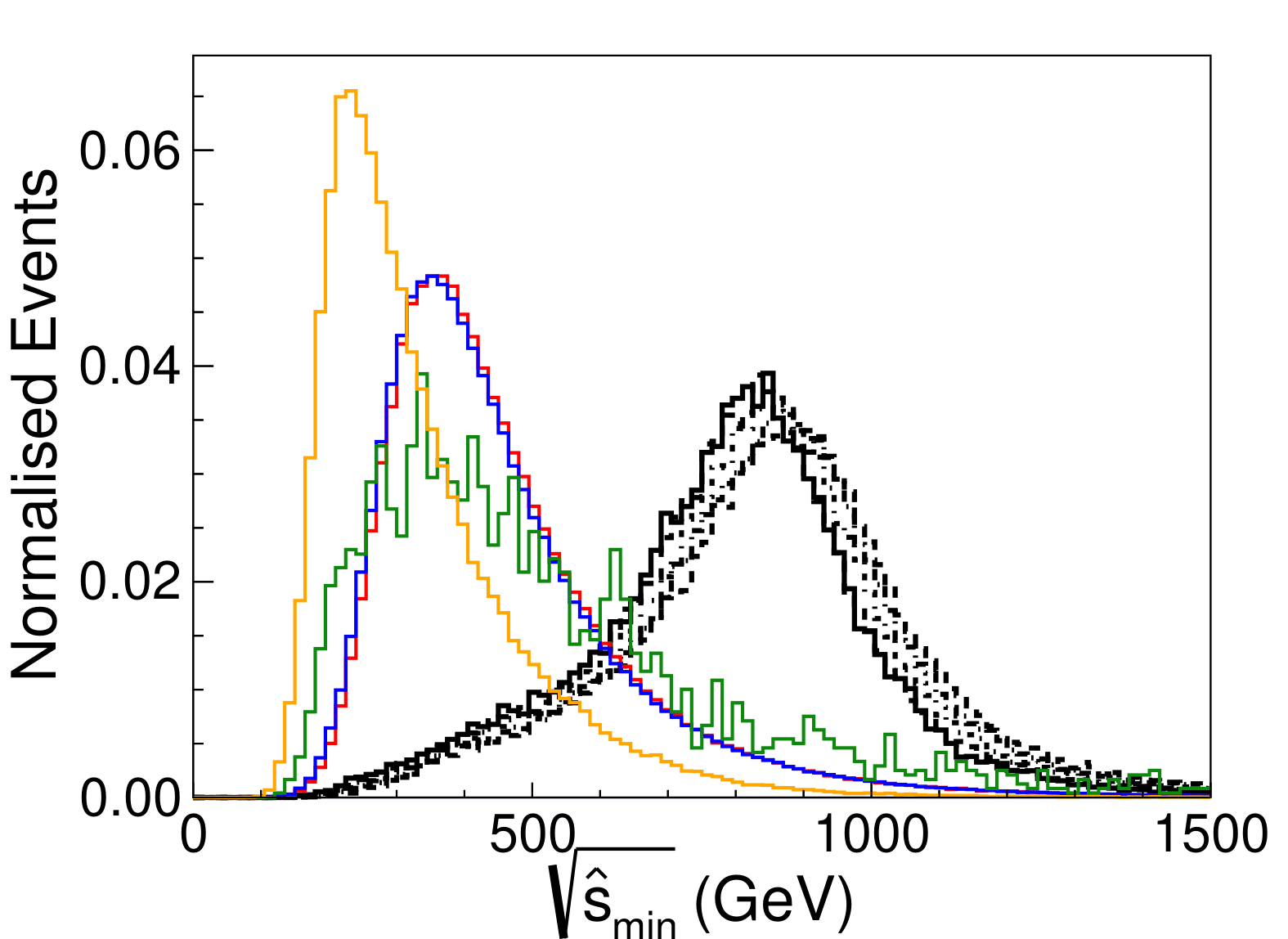}}
\subfigure[NoL]{
\includegraphics[height = 4.5 cm, width = 8.0 cm]{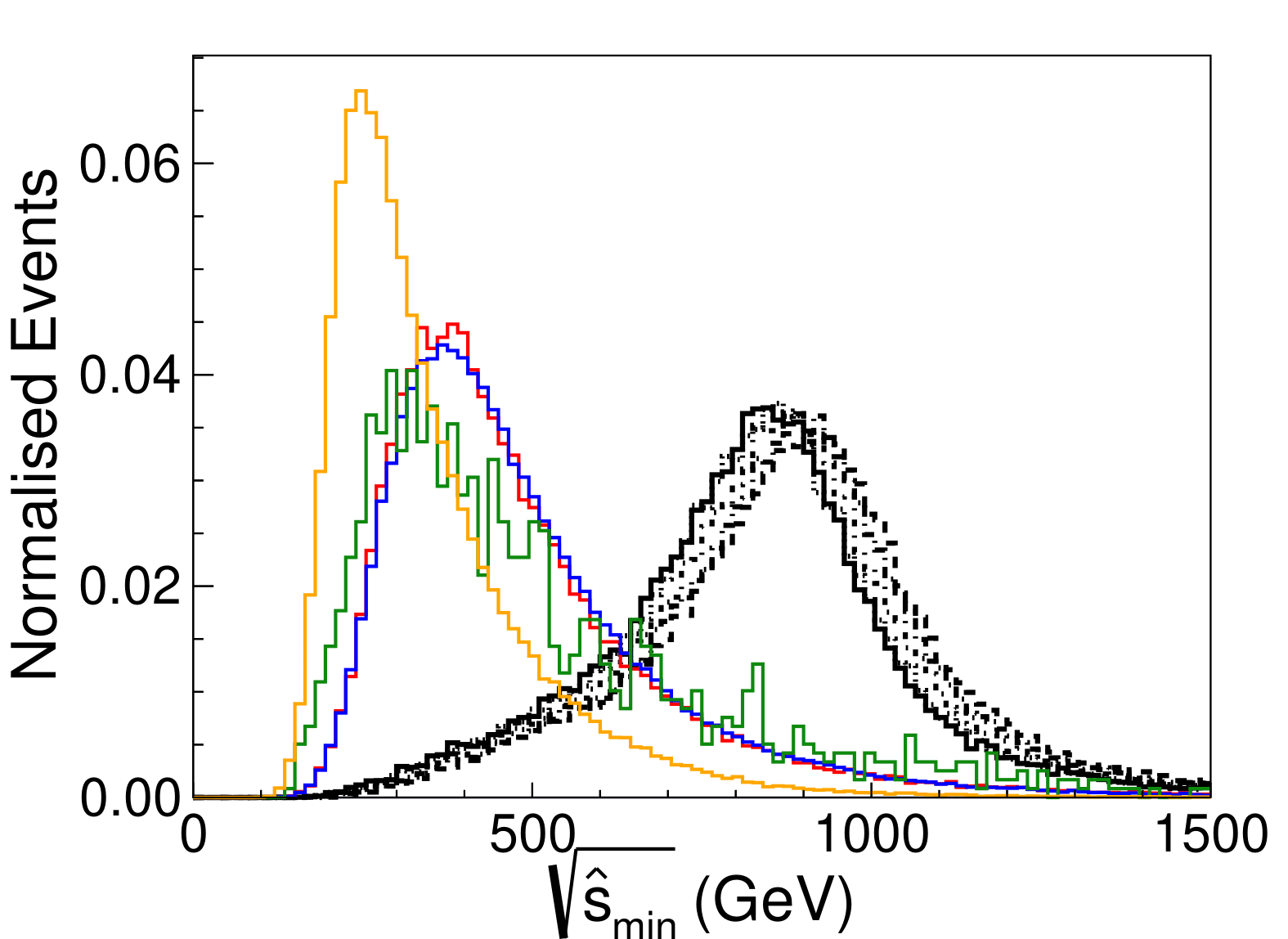}} 
} 
 \hspace{0.01\textwidth}
  \centering
  \includegraphics[width=0.7\textwidth]{Legend_new.png}
\caption{ Distributions of some kinematic variables: (a,b) $\Delta R$ between the decay products of $A$, (c,d) $\sqrt{\hat{s}_{min}}$, for SL and NoL channels respectively.}
\label{fig:features-2}
\end{figure}

Thus the baseline selection criterion are mainly used to ensure the presence of correct final state particles in signal and background events. As can be seen from Table \ref{tab:Yields}, at 
an integrated luminosity ${\cal L}\,=\,3000\,{\rm fb^{-1}}$, after applying the cuts C0-C4, signal to background ratio for each benchmark turns out to be small. Thus achieving a good signal significance becomes quite challenging if we only use C0-C4. However, a few kinematic variables seem to have better discriminative power to classify signal events over background as shown in Fig.\ref{fig:features-1} and Fig.\ref{fig:features-2}. Let us provide a brief description of these variables and impose appropriate cuts (C5-C9) on them to maximize the signal significance.

\begin{itemize}
  
\item[C5:] We have depicted the normalized distributions of the transverse momentum of the leading $b$-jet ($p_T^{b_1}$) for all benchmarks and dominant backgrounds for SL and NoL channels in Fig.\ref{fig:features-1}(a) and \ref{fig:features-1}(b) respectively. Since the $b$-jets originate from the decay of a heavy particle $S_R$ having mass $\sim 870-950$ GeV, the corresponding distributions of $p_T^{b_1}$ for the signal are harder than that of the backgrounds. Thus we demand $p_T^{b_1} > 200$ GeV to destroy the backgrounds over the signals.

\item[C6:] Similarly, for the sub leading $b$-jet, the distributions of $p_T^{b_2}$ for the signals and backgrounds are plotted in Fig.\ref{fig:features-1}(c) and \ref{fig:features-1}(d) respectively for SL and NoL channels. The nature of the distributions can be explained by applying the same logic given in C5. In this case, an efficient discrimination of signals and backgrounds would require $p_T^{b_2} > 100$ GeV.  

\item[C7:] The normalized distributions of $\Delta R_{b_1 b_2}$ for signal and backgrounds in the SL and NoL channels are drawn in Fig.\ref{fig:features-1}(e) and \ref{fig:features-1}(f). In both channels two $b$-jets originate from the massive particle $S_I$. As a result $S_I$ is not boosted enough to keep it's decay products collimated. Thus the distributions of $\Delta R_{b_1 b_2}$ for signal peak at a higher side than the backgrounds in both channels. Therefore to achieve maximum significance we impose a lower cut : $\Delta R_{b_1 b_2} > 2.0$.   

\item[C8:] Another important variable with a reasonable distinguishing power between the signal and backgrounds is $\Delta R_{\ell \tau_h}$ ($\Delta R_{\tau_{h_1} \tau_{h_2}}$) for SL (NoL) channel. Corresponding normalized distributions are shown in Fig.\ref{fig:features-2}(a) and \ref{fig:features-2}(b) respectively for SL and NoL channels. The visible decay products of $\tau^+ \tau^-$ in the semi-leptonic and fully hadronic decay modes originate from a lighter pseudoscalar with mass $\sim 66-147$ GeV. Thus the  final state lepton and $\tau$-jet (two $\tau$-jets) in SL (NoL) channel become collimated, thereby setting $\Delta R_{\ell \tau_h}$ ($\Delta R_{\tau_{h_1} \tau_{h_2}}$) to a smaller value for signal compared to the backgrounds. Thus we apply an upper cut : $\Delta R_{\ell \tau_h}$ ($\Delta R_{\tau_{h_1} \tau_{h_2}}$) $< 1.8 $ to banish the backgrounds.

\item[C9:] Finally we use {\em minimum parton level centre-of-mass energy} ($\sqrt{\hat{s}_{min}}$) which has highest degree of discerning power between the signal and backgrounds. Basically this is a global inclusive variable for determining the mass scale of any new physics in presence of missing energy at the final states. The normalized distributions for both channels and for signal and backgrounds are depicted in Fig.\ref{fig:features-2}(c) and \ref{fig:features-2}(d). Since this variable plays a significant role in wiping out the backgrounds, the signal significance is expected to be sensitive to it. Thus instead of giving a fixed cut on this variable, we try to tune $\sqrt{\hat{s}_{min}}$ over a suitable range to maximize the significance. Thus we do not include this cut (C9) in the cut-flow Table \ref{tab:Yields}. Table \ref{tab:smincut} shows the variation of signal significance with $\sqrt{\hat{s}_{min}}$. For example for BP2, the significance increases by 20$\%$ (14.8$\%$) for SL (NoL) channel after using this variable. 

\end{itemize}

\begin{table}[htpb!]
  \begin{center}
    {\footnotesize
      \centering
      \setlength{\tabcolsep}{0.7em} 
                {\renewcommand{\arraystretch}{1.2}
                  \begin{tabular}{|l|c|c|c|c|c|c|}
                    \hline
                    \multirow{2}{*}{Processes}                                 & Events                    &    \multicolumn{5}{c|}{Events after cuts}            \\ \cline{3-7}
                                                                               & produced                  &    C0-C4    &    C5    &    C6   &    C7   &    C8     \\
                    \hline
                    \multicolumn{7}{|c|}{\texttt{Signal Benchmarks}} \\
                    \hline
                    \multirow{2}{*}{BP1}                                       & \multirow{2}{*}{$129300$} &   $3371$  &  $2763$  & $2377$  & $2221$  & $1842$    \\
                                                                               &                           &   $4097$  &  $3332$  & $2791$  & $2564$  & $1994$    \\
                    \hline
                    \multirow{2}{*}{BP2}                                       & \multirow{2}{*}{$128700$} &   $3892$  &  $3171$  & $2714$  & $2518$  & $1924$    \\
                                                                               &                           &   $4604$  &  $3750$  & $3134$  & $2870$  & $2036$    \\
                    \hline
                    \multirow{2}{*}{BP3}                                       & \multirow{2}{*}{$102600$} &   $3658$  &  $3024$  & $2586$  & $2389$  & $1608$    \\
                                                                               &                           &   $4184$  &  $3443$  & $2889$  & $2640$  & $1649$    \\
                    \hline
                    \multirow{2}{*}{BP4}                                       & \multirow{2}{*}{$62700$}  &   $2520$  &  $2095$  & $1793$  & $1652$  & $974$     \\
                                                                               &                           &   $2764$  &  $2288$  & $1931$  & $1762$  & $971$     \\
                    \hline
                    \multirow{2}{*}{BP5}                                       & \multirow{2}{*}{$20400$}  &   $905$   &  $756$   & $645$   & $593$   & $293$     \\
                                                                               &                           &   $977$   &  $812$   & $682$   & $622$   & $282$     \\
                    \hline
                    \multicolumn{7}{|c|}{\texttt{Standard Model Backgrounds with Major Contributions}} \\
                    \hline
                    \multirow{2}{*}{$t\bar{t}\,\to\,2\ell\,+\,jets$} &   \multirow{2}{*}{$3.23\times 10^8$}   & $7343240$ & $564720$ & $287951$ & $261605$ & $54530$   \\
                                                                     &                                        & $723852$  &  $66376$ & $33086$  &  $29546$ & $6348$    \\
                    \hline
                    \multirow{2}{*}{$t\bar{t}\,\to\,1\ell\,+\,jets$} &   \multirow{2}{*}{$1.31\times 10^9$}   & $4773602$ & $469033$ & $229027$ & $187641$ & $52153$   \\
                                                                     &                                        & $1119938$ & $125423$ & $59333$  & $47860$  & $12950$   \\
                    \hline
                    \multirow{2}{*}{$tW$}                            &   \multirow{2}{*}{$1.03\times 10^8$}   & $2658814$ & $126566$ & $64578$  & $59989$  & $12302$   \\
                                                                     &                                        & $234436$  & $13484$  & $7001$   & $6378$   & $1368$    \\ 
                    \hline
                    \multirow{2}{*}{$t\bar{t}Z\,\to\,\ell^+\ell^-\,+\,jets$}   & \multirow{2}{*}{$720000$}    & $12956$   & $2285$   & $1171$   &   $930$  &  $480$    \\
                                                                               &                              & $7637$    & $1405$   &  $694$   &    $541$ &  $362$    \\
                    \hline
                    \multirow{2}{*}{$WZ\,\to\,2\ell\,2q$}            & \multirow{2}{*}{$1.35\times 10^7$}     & $3550$    & $687$    &  $283$   & $223$    &  $136$    \\
                                                                     &                                        & $3130$    & $556$    &  $229$   & $169$    &  $131$    \\
                    \hline
                    \multirow{2}{*}{$t\bar{t}W\,\to\,\ell \nu\,+\,jets$}       & \multirow{2}{*}{$762000$}    & $7703$    & $1321$   & $635$    &   $467$  &  $128$    \\
                                                                               &                              & $1144$    &  $213$   & $100$    &    $73$  &   $22$    \\
                    \hline
                    
      \hline
    \end{tabular}}}
    \end{center}
    \caption{Event yields of the signal and SM background processes after the baseline selection (C0-C4) and after each successive selection cuts (C5-C8) of the cut based analysis at the 14\,TeV LHC for ${\cal L}\,=\,3000\,{\rm fb^{-1}}$. Each row is divided into two subrows that contain the information of the SL and NoL channels, respectively.}
    \label{tab:Yields}
\end{table}

In Table \ref{tab:Yields} we tabulate the signal (BP1-BP5) and background yields at integrated luminosity 3000 fb$^{-1}$ after imposing the baseline selection cuts (C0-C4) and successive cuts on relevant kinematic variables (C5-C9). Looking at the signal significances in Table \ref{tab:smincut}, one can conclude that NoL channel turns out to be the most promising among the two channels at 14 TeV HL-LHC. In the same table we also turn on $5 \%$ systematic uncertainty and evaluate the reduced signal significance. Due to huge background contribution, a $5\%$ systematic uncertainty on background affects the signal significance by a large margin. So, we proceed to perform more sophisticated multivariate analysis to achieve a better signal significance.

\begin{table}[htpb!]
  \begin{center}
    {\footnotesize
      \centering
      \setlength{\tabcolsep}{0.7em} 
                {\renewcommand{\arraystretch}{1.2}
                  \begin{tabular}{|l|c|c|c|c|c|}
                    \hline
                    \multirow{2}{*}{Processes}    & Cut on                     &       \multicolumn{2}{c|}{Remaining events}     &  \multicolumn{2}{c|}{Significance}      \\ \cline{3-6}
                                                  & $\sqrt{\hat{s}_{min}}$     &          Signal        &      Background        &  $\theta\,=\,0\%$  &  $\theta\,=\,5\%$  \\
                    \hline
                    \multirow{2}{*}{BP1}          & $718$                      &          $1568$        &        $60639$         &      $6.37$        &      $0.51$        \\
                                                  & $682$                      &          $1835$        &        $13639$         &      $15.7$        &      $2.65$        \\
                    \hline
                    \multirow{2}{*}{BP2}          & $718$                      &          $1658$        &        $60640$         &      $6.73$        &      $0.54$        \\
                                                  & $694$                      &          $1867$        &        $13316$         &      $16.2$        &      $2.76$        \\
                    \hline
                    \multirow{2}{*}{BP3}          & $742$                      &          $1388$        &        $55728$         &      $5.88$        &      $0.49$        \\
                                                  & $742$                      &          $1463$        &        $11910$         &      $13.4$        &      $2.42$        \\
                    \hline
                    \multirow{2}{*}{BP4}          & $766$                      &          $834$         &        $51065$         &      $3.69$        &      $0.32$        \\
                                                  & $742$                      &          $883$         &        $11910$         &      $8.09$        &      $1.46$        \\
                    \hline
                    \multirow{2}{*}{BP5}          & $790$                      &          $250$         &        $46768$         &      $1.15$        &      $0.11$        \\
                                                  & $742$                      &          $259$         &        $11910$         &      $2.37$        &      $0.43$        \\
                    
                    \hline
    \end{tabular}}}
    \end{center}
    \caption{Best cut on $\sqrt{\hat{s}_{min}}$ and corresponding signal and background yields for the five signal benchmark points. Last two columns show the signal significance values at ${\cal L}\,=\,3000\,{\rm fb^{-1}}$ with and without a systematic uncertainty $(\theta)$ of 0 and 5$\%$, respectively.}
    \label{tab:smincut}
\end{table}

\subsection{Multivariate analysis}

We use deep neural network (DNN) \cite{lecun2015deep} to perform the multivariate analysis (MVA). We follow a supervised learning technique to do a binary-class classification. The basic work flow of a DNN is the following:

A DNN has more than one hidden layer with multiple nodes or neurons fully connected to the nodes of the consecutive layers via different weights. The input to each node of $n^{th}$ is the linear superposition of the outputs of all the nodes in layer $(n-1)$. A nonlinear activation function is then applied on each node of layer $n$. The final layer of a network is the output layer and the output is estimated in terms of probability which is a function of all the weights of the network. The difference between the true output and the predicted one is referred as the loss function. The loss function is then minimized using stochastic gradient descent method to extract the best values of the weights. Those optimised weights represent a suitable nonlinear boundary on the plane of the input features that can classify the signal and background events.

For all the five signal benchmarks, we train different networks of the same architecture and using the same set of input variables. We use a residual network (ResNet) like architecture rather than a simple feed forward network to perform this study. A ResNet has shortcut connections between multiple layers to make sure the gradient for minimization do not vanishes. This kind of structure is better to train a deeper network. One can see the detailed concept of ResNet in Reference \cite{DBLP:journals/corr/HeZRS15}. We use $80\%$ of the whole dataset {\it i.e.} signal and background combined, as the training set and to evaluate the performance of corresponding models, we keep the remaining set referred as the test dataset. The input variables used for training are described in Table \ref{tab:features}.

\begingroup \setlength{\tabcolsep}{7pt}
\renewcommand{\arraystretch}{1}
\begin{table}[htpb!]
  \begin{center}
    \footnotesize\setlength{\extrarowheight}{1pt}
    \begin{tabular}{|l|c|c|c|}
      \hline
      \multirow{2}{*}{No.} & \multicolumn{2}{|c|}{Variables}                                             & Description                                     \\ \cline{2-3}
                           & \texttt{SL}                   &             \texttt{NoL}                    & SL\,(NoL)                                       \\
      \hline
      1  & \multicolumn{2}{c|}{\texttt{$p_T^{b_1}$}}                                  & $p_T$ of leading $b$-jet                    \\
      2  & \multicolumn{2}{c|}{\texttt{$p_T^{b_2}$}}                                  & $p_T$ of sub-leading $b$-jet                     \\
      3  & \multicolumn{2}{c|}{\texttt{$|\eta^{b_1}|$}}                               & $|\eta|$ of leading $b$-jet                    \\
      4  & \multicolumn{2}{c|}{\texttt{$|\eta^{b_2}|$}}                               & $|\eta|$ of sub-leading $b$-jet                     \\
      5  & \multicolumn{2}{c|}{\texttt{$\met$}}                                       & Missing transverse energy                    \\ \cline{2-3}
      6  & \texttt{$p_T^{\tau_h}$}       & \texttt{$p_T^{\tau_h^1}$}                   & $p_T$ of leading $\tau$-jet                    \\
      7  & \texttt{$|\eta^{\tau_h}|$}    & \texttt{$|\eta^{\tau_h^1}|$}                & $|\eta|$ of leading $\tau$-jet                    \\
      8  & \texttt{$p_T^{\ell}$}         & \texttt{$p_T^{\tau_h^2}$}                   & $p_T$ of lepton\,(sub-leading $\tau$-jet)                    \\
      9  & \texttt{$|\eta^{\ell}|$}      & \texttt{$|\eta^{\tau_h^2}|$}                & $|\eta|$ of lepton\,(sub-leading $\tau$-jet)                    \\ 
      10 & \texttt{$\Delta R_{\ell, \tau_h}$} & \texttt{$\Delta R_{\tau_h^1, \tau_h^2}$} & $\Delta R$ between lepton-$\tau_h$ ($\tau_h^1$-$\tau_h^2$) coming from $A$ \\
      11 & \texttt{$\Delta \phi_{\ell,\, \met}$} & \texttt{$\Delta \phi_{\tau_h^2, \met}$} & $|\Delta \phi|$ between lepton-$\met$ ($\tau_h^2$-$\met$)                   \\
      12 & \texttt{$\Delta R_{\tau_h, A}$} & --                                        & $\Delta R$ between $\tau_h$ and reconstructed $A$ \\
      13 & \texttt{$\Delta R_{\tau_h, ssr}$} & \texttt{$\Delta R_{\tau_h^1, ssr}$}     & $\Delta R$ between $\tau_h$\,($\tau_h^1$) and reconstructed $ssr$ {\it i.e.} $b\bar{b}$ \\
      14 & \texttt{$\Delta R_{\ell, \tau_h} \times p_T^A$} & \texttt{$\Delta R_{\tau_h^1, \tau_h^2} \times p_T^A$} & No. 10 $\times~ p_T^A$ \\ \cline{2-3}
      15 & \multicolumn{2}{c|}{\texttt{$\Delta R_{b_1, b_2}$}}                         & $\Delta R$ between leading and sub-leading $b$-jet \\
      16 & \multicolumn{2}{c|}{\texttt{$\Delta R_{b_1, b_2} \times p_T^{ssr}$}}        & No. 15 $\times ~ p_T{b1,b2}$  \\ \cline{2-3}
      17 & \texttt{$\Delta R_{\ell, b_1}$} & \texttt{$\Delta R_{\tau_h^1, b_1}$}       & $\Delta R$ between lepton\,($\tau_h^1$) and leading $b$-jet \\
      18 & \texttt{$\Delta R_{\ell, b_2}$} & \texttt{$\Delta R_{\tau_h^1, b_2}$}       & $\Delta R$ between lepton\,($\tau_h^1$) and sub-leading $b$-jet \\
      19 & \texttt{$\Delta R_{\tau_h, b_1}$} & \texttt{$\Delta R_{\tau_h^2, b_1}$}       & $\Delta R$ between $\tau_h$\,($\tau_h^2$) and leading $b$-jet \\
      20 & \texttt{$\Delta R_{\tau_h, b_2}$} & \texttt{$\Delta R_{\tau_h^2, b_2}$}       & $\Delta R$ between $\tau_h$\,($\tau_h^2$) and sub-leading $b$-jet \\ \cline{2-3}
      21 & \multicolumn{2}{c|}{\texttt{$\Delta \phi_{b_1,\,\met}$}}                     & $|\Delta \phi|$ between leading $b$-jet and $\met$ \\
      22 & \multicolumn{2}{c|}{\texttt{$\Delta \phi_{b_2,\,\met}$}}                     & $|\Delta \phi|$ between sub-leading $b$-jet and $\met$ \\
      23 & \multicolumn{2}{c|}{\texttt{$\Delta R_{b_1, A}$}}                            & $\Delta R$ between leading $b$-jet and reconstructed $A$ \\
      24 & \multicolumn{2}{c|}{\texttt{$\Delta R_{min}^{jets}$}}                        & Minimum $\Delta R$ between all jets \\
      25 & \multicolumn{2}{c|}{\texttt{$\sqrt{\hat{s}_{min}}$}}                         & Minimum parton-level centre-of-mass energy \\
      26 & \multicolumn{2}{c|}{\texttt{$n-Jets$}}                                       & Number of jets \\  
      \hline
    \end{tabular}
    \footnotesize
  \end{center}
  \caption{ Input variables used for DNN.}
  \label{tab:features}
\end{table}

We try to choose the important input features by estimating the F-score using permutation invariance \cite{Breiman2001} for each analysis channel and signal benchmark. The input layer of the DNN is equipped with these 26(25) features described in Table \ref{tab:features}. Then the ResNet structure follows an initial hidden layer with $n_0=512$ nodes connected to five shortcut hops ($i$) each consists of two layers with equal number of nodes decreasing by a fraction of $0.5$ {\it i.e.} $n_i = 0.5 \times n_{i-1}$. Finally, two more hidden layers with 8 and 4 nodes and, then the output layer having two nodes represent signal and background. The activation function used in every hidden layer is the rectified linear-unit (``relu'') function and ``sigmoid'' function is used at the output to get the classification probability for signal and background events. Rest of the model parameters are described in Table \ref{tab:dnnparams}.

\begin{table}[htpb!]
  \begin{center}
    \footnotesize\setlength{\extrarowheight}{1pt}     
    \begin{tabular}{|l|c|c|}
      \hline
      Parameters                 & Description                                                   & Values/Choices  \\
      \hline
      \texttt{loss\_function}    & Function to be minimised to get optimum model parameters      & $binary\_crossentropy$ \\
      \texttt{optimiser}         & Perform gradient descent and back propagation                 & $Adam$  \\  
      \texttt{eta}               & Learning rate                                                 & $0.001$ \\
      \texttt{batch\_len}        & Number of events in each mini batch                           & $5000$  \\ 
      \texttt{batch\_norm}       & Normalisation of activation output                            & $True$  \\
      \texttt{dropout}           & Fraction of random drop in number of nodes                    & $20\%$  \\
      \texttt{L2-Regularizer}    & Regularize loss to prevent over-fitting                        & $0.0001$ \\      
      \hline
    \end{tabular}
    \footnotesize
  \end{center}
  \caption{Details of the DNN parameters.}
  \label{tab:dnnparams}
\end{table}

After training, we check the performances of respective DNN models on the test dataset. Figure \ref{fig:ROCs} show the receiver operating characteristic (ROC) curves and corresponding area under the curve {\it i.e.} AUC values for BP1. The degree of performance of the MVA techniques increases with increasing AUC and the other benchmarks exhibit similar nature. 


\begin{figure}[htpb!]
\centering
\subfigure[SL (BP1)]{
\includegraphics[scale=0.32]{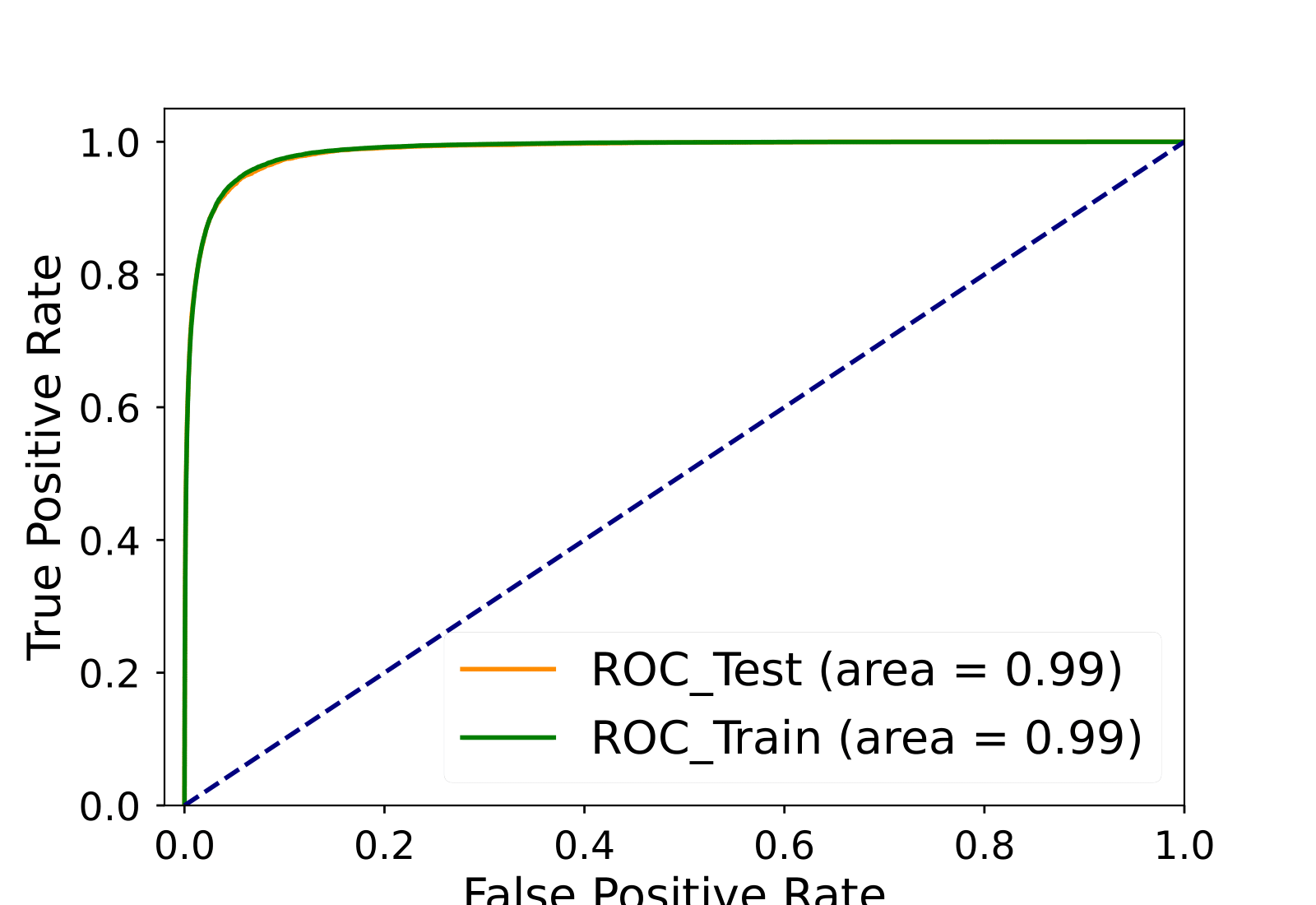}}
\subfigure[NoL (BP1)]{
\includegraphics[scale=0.32]{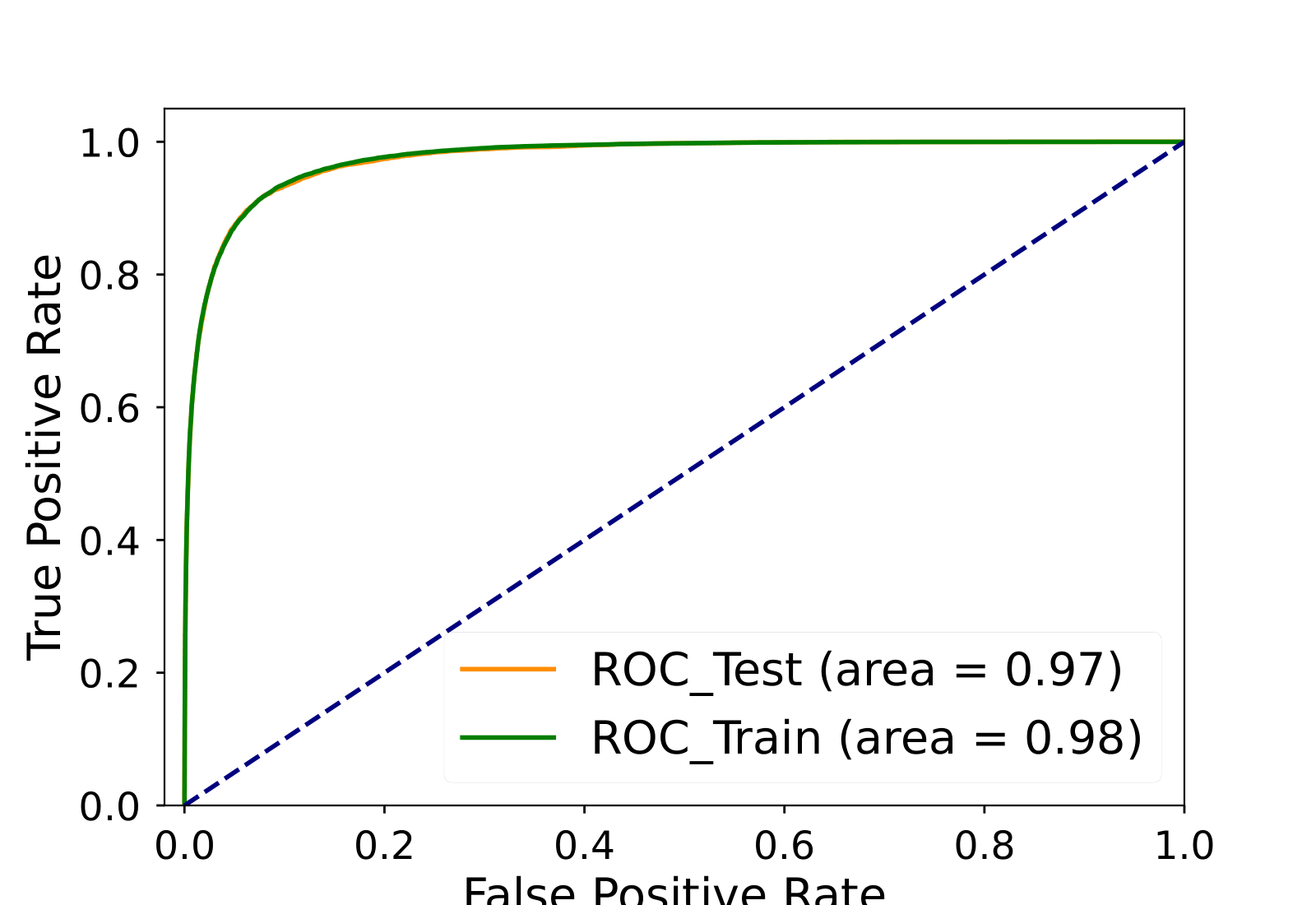}}
\caption{ROCs of the DNN trained for $M_A\,=\,66.39$ GeV in (a) SL and, (b) NoL channels. True positive rate and false positive rate describe the signal efficiency and background efficiency, respectively.}
\label{fig:ROCs}
\end{figure}

The models are trained in a stochastic approach and therefore, with increasing the number of iteration, the loss is expected to decrease because the network tries to learn the nature of signal and background from the distributions of the input features. We observe similar behavior for the loss and ROC for both train and test data which indicate the presence of negligible over-training. Based on that, we proceed to use respective models to evaluate the significance of the signal benchmarks. We also consider a $5\%$ linear-in-background systematic uncertainty on the background contribution to see the effect in signal significance values.

\begin{table}[h!]
  \begin{center}
    {\footnotesize
      \centering
      \setlength{\tabcolsep}{0.7em} 
                {\renewcommand{\arraystretch}{1.2}
                  \begin{tabular}{|l|c|c|c|c|c|}
                    \hline
                    \multirow{2}{*}{Processes}    & Cut on                     &       \multicolumn{2}{c|}{Remaining events}     &  \multicolumn{2}{c|}{Significance}      \\ \cline{3-6}
                                                  & DNN response               &          Signal        &      Background        &  $\theta\,=\,0\%$  &  $\theta\,=\,5\%$  \\
                    \hline
                    \multirow{2}{*}{BP1}          & $0.91$                     &          $1201$        &        $6228$          &      $15.2$        &      $3.74$        \\
                                                  & $0.67$                     &          $2470$        &        $12806$         &      $21.8$        &      $3.80$        \\
                    \hline
                    \multirow{2}{*}{BP2}          & $0.89$                     &          $1442$        &        $7852$          &      $16,3$        &      $3.58$        \\
                                                  & $0.68$                     &          $2663$        &        $13526$         &      $22.9$        &      $3.88$        \\
                    \hline
                    \multirow{2}{*}{BP3}          & $0.95$                     &          $1400$        &        $9084$          &      $14.7$        &      $3.02$        \\
                                                  & $0.67$                     &          $2137$        &        $11328$         &      $20.1$        &      $3.71$        \\
                    \hline
                    \multirow{2}{*}{BP4}          & $0.88$                     &          $641$         &        $3926$          &      $10.2$        &      $3.11$        \\
                                                  & $0.88$                     &          $1194$        &        $6643$          &      $14.6$        &      $3.49$        \\
                    \hline
                    \multirow{2}{*}{BP5}          & $0.97$                     &          $167$         &        $3542$          &      $2.80$        &      $0.89$        \\
                                                  & $0.91$                     &          $264$         &        $2431$          &      $5.36$        &      $2.02$        \\
                    
                    \hline
    \end{tabular}}}
    \end{center}
    \caption{ Best cut on DNN response and corresponding signal and background yields for the five signal benchmark points. Last two columns show the signal significance values at ${\cal L}\,=\,3000\,{\rm fb^{-1}}$ with and without a systematic uncertainty $(\theta)$ of 0 and 5$\%$, respectively.}
    \label{tab:dnncut}
\end{table}

Table \ref{tab:dnncut} demonstrates the best possible cut on the DNN responses for every signal benchmarks keeping $B \geq 5 \times S$ \cite{Cowan:2010js} to make sure of everything remains in the asymptotic regime. The last column of Table \ref{tab:dnncut} shows the effect of a $5\%$ linear-in-background systematic uncertainty on the signal significance. It is observed that the DNN performs better than the cut based method. For instance, the statistical significance for BP5 improves by a factor $\gtrsim$ 2 upon switching to DNN. In absence of systematic uncertainties, this makes it possible to discover a pseudoscalar of mass 147 GeV in BP5 at 5$\sigma$.

\section{Summary and conclusions}\label{conclusions}

The recently reported discrepancy between the measured value of $M_W$
and its SM prediction has stirred up fresh hopes of having observed BSM phenomena. At the same time, the lingering excess in the muon anomalous magnetic moment of the muon has also opened door to model building using BSM physics. In thus study, we have proposed a solution to the twin anomalies in the framework comprising both color-singlet as well as color-octet scalars. More precisely, the well-known Type-X 2HDM was augmented with the color octet isodoublet. Particular emphasis has been laid on the role of the colored scalars in this context. That is, a virtual contribution of the colored scalars to the oblique parameters aids to uplift the $W$-mass to the observed value. At the same time, two-loop Barr-Zee contributions induced  by the colored scalars extend the parameter region compatible with muon $g-2$
\emph{w.r.t.} what is seen for the pure Type-X 2HDM.

We have proposed the $p p \to S_R \to S_I A \to b \bar{b} \tau^+ \tau^-$ signal in this work to look for the various scalars involved, both colorless as well as colored. The final ensuing $b\bar{b}\tau\tau$ final state is attractive from the perspective of collider experiments. This signal has been analysed at the 14 TeV LHC using both cut-based as well as multivariate techniques, in particular, deep neural networks. We have found that the observability of the framework appreciably improves upon incorporating  DNN. One must also note that the effect of systematics is also quite high in the statistical significances due to high amount of background contamination. Several sources of systematics are not taken care of, such as: jet to $\tau_h$ fake, lepton to jet fake, pdf error, several normalised and shape based scale factors templates etc. By proper implementation of all the experimental details, such signal topologies have the potential to unravel the presence of both colorless as well as color octer scalars at the HL-LHC.

\acknowledgements IC acknowledges support from Department of Science and Technology, Govt. of India, under grant number IFA18-PH214 (INSPIRE Faculty Award). NC acknowledges support from Department of Science and Technology, Govt. of India, under grant number IFA19-PH237 (INSPIRE Faculty Award).
The authors also acknowledge support of the computing
facilities of Indian Association for the Cultivation Science, Saha Institute of Nuclear Physics and Indian Institute of Technology Kanpur.

\section{Appendix}\label{appendix}

\subsection{Yukawa scale factors}

\begin{table}[htpb!]
\centering
\begin{tabular}{ |c c c c c c c c c| } 
\hline
$\xi^h_e$ & $\xi^h_\mu$ & $\xi^h_\tau$ & $\xi^H_e$ & $\xi^H_\mu$ & $\xi^H_\tau$ & $\xi^A_e$ & $\xi^A_\mu$ & $\xi^A_\tau$ \\ \hline
 $-\frac{\text{sin}\a}{\text{cos}\b}$
& $-\frac{\text{sin}\a}{\text{cos}\b}$
& $-\frac{\text{sin}\a}{\text{cos}\b}$
& $\frac{\text{cos}\a}{\text{cos}\b}$
& $\frac{\text{cos}\a}{\text{cos}\b}$
& $\frac{\text{cos}\a}{\text{cos}\b}$
& tan$\beta$ & tan$\beta$ & tan$\beta$ \\ \hline
\end{tabular}
\caption{Various Yukawa scale factors for the lepton-specific case.}
\label{tab:xi}
\end{table}

\subsection{Functions in the two-loop BZ amplitudes }
\besub
\bea
\mathcal{F}^{(2)} (z) = \frac{1}{2} \int_{0}^{1} dx \frac{x(1-x)}{z-x(1-x)} ~{\rm ln} \left(\frac{z}{x(1-x)}\right), \\
\mathcal{G}(z^a,z^b,x) = \frac{{\rm ln} \left(\frac{z^a x + z^b (1-x)}{x(1-x)}\right)}{x(1-x) - z^a x - z^b (1-x)}.
\eea
\eesub

\bibliography{ref} 
\end{document}